\documentclass[a4paper]{article}
\usepackage{amsfonts}
\usepackage{mathrsfs}
\usepackage{amsmath}
\usepackage{amssymb}
\usepackage{xcolor}
\usepackage{amsfonts}
\usepackage{graphicx}
\usepackage{subfigure}
\usepackage{cite}
\usepackage{hyperref}
\usepackage{algorithm}
\usepackage{algorithmicx}
\usepackage[letterpaper,top=2cm,bottom=2cm,left=3cm,right=3cm,marginparwidth=1.75cm]{geometry}

\usepackage{multirow}   
\usepackage{array}      

\newcommand{\argmin}{\operatorname*{arg\,min}}

\title{Fast algorithms enabling optimization and deep learning
for photoacoustic tomography in a circular detection geometry}

\author{Andreas Hauptmann\thanks{Research Unit of Mathematical Sciences, University of Oulu, Oulu, Finland and the Department of Computer Science, University College London, London, U.K.}, Leonid Kunyansky\thanks{Department of Mathematics, University of Arizona, Tucson, USA}, and Jenni Poimala\thanks{Department of Technical Physics, University of Eastern Finland, Kuopio, Finland}}

\begin{document}

\maketitle

\begin{abstract}
The inverse source problem arising in photoacoustic tomography and in several other coupled-physics modalities is frequently solved by iterative algorithms. Such algorithms are based on the minimization of a certain cost functional. In addition, novel deep learning techniques are currently being investigated to further improve such optimization approaches. All such methods require multiple applications of the operator defining the forward problem, and of its adjoint.
In this paper, we present new asymptotically fast algorithms for numerical evaluation of the forward and adjoint operators, applicable in
the circular acquisition geometry. For an $(n \times n)$ image, our algorithms compute these operators in $\mathcal{O}(n^2 \log n)$ floating point operations. We demonstrate the performance of our algorithms in numerical simulations, where they are used as an integral part of several iterative image reconstruction techniques: classic variational methods, such as non-negative least squares and total variation regularized least squares, as well as deep learning methods, such as learned primal dual.
A Python implementation of our algorithms and computational examples is  available to the general public \thanks{Codes available at \url{https://github.com/leonidak/torch-TAT}}.
\end{abstract}

\section{Introduction}

Image reconstruction in a large number of novel coupled-physics (or hybrid)
imaging modalities involves a certain wave operator. For example, two of the
most developed hybrid modalities are photoacoustic and thermoacoustic tomography
(PAT\cite{Kruger95,Oraev94} and TAT\cite{Kruger99}). In PAT and TAT such a
wave operator describes propagation of the acoustic wave generated by an
instantaneous thermoelastic expansion of tissues induced by a short pulse of
infrared or microwave radiation. The acoustic wave in this case is a solution
of the Cauchy problem for the wave equation in $\mathbb{R}
^{d}$:
\begin{align}
\frac{\partial^{2}}{\partial t^{2}}p(t,x)  &  =c^{2}\Delta p(t,x),\quad
t\geq0,\quad x\in\mathbb{R}^{d},\label{E:wave-eq}\\
p(0,x)  &  =f(x),\quad\frac{\partial}{\partial t}p(0,x)=0, \label{E:Cauchy}
\end{align}
where $p(t,x)$ is the excess acoustic pressure, $c$ is the known and constant
speed of sound, and function $f(x)$ represents the initial pressure due to
thermoelastic expansion. The initial pressure $f$  is assumed to be compactly supported
within a region $\Omega_{0}$ that is a subset of a convex domain $\Omega$ with
a smooth boundary $\partial\Omega$. The trace $g(t,z)$ of the pressure is
measured during the time interval $t\in(0,T]$ by a set of transducers that cover
a subset $\Gamma\subset\partial\Omega$ of the boundary:
\[
g(t,z)=p(t,z),\quad(t,z)\in(0,T]\times\Gamma.
\]
This defines the wave operator $\mathcal{A}$:
\[
\mathcal{A}:f\mapsto g.
\]

Questions about the injectivity and invertibility of $\mathcal{A}$
\cite{Finch04,Finch07,KuKuEncycl,StUhl}, as well as the design of inversion
algorithms have received a lot of attention over the last two decades
\cite{Norton1,MXW1,Wang-universal,Finch04,Finch07,KunyanskyIP07,Nguyen,Kun-cyl,Pala,Haltm-convex,Do-Kun18,EllerKun}. In the simplest cases (for example, when $\Gamma$ is a circle or a sphere
completely surrounding the support of $f$) the operator $\mathcal{A}$ is
stably invertible, there are explicit formulas for $\mathcal{A}^{-1}
$\cite{Norton1,MXW1,Wang-universal,Finch04,Finch07,KunyanskyIP07}, and
efficient computational algorithms have been developed
\cite{Norton1,MXW1,Wang-universal,Kun-cyl,Haltm_circ,Haltm_cyl,DigitalTwin2025} to compute
$\mathcal{A}^{-1}g$ and, thus, to reconstruct the sought initial condition
$f.$ However, in more complex cases involving limited-view geometries or the measurements affected by high level of noise,
approximations to $f(x)$ are often computed with variational methods, which seek to reconstruct $f(x)$ as the minimizer of a cost functional by applying suitable optimization algorithms.
Computing such solutions usually requires multiple numerical evaluations of the forward operator
$\mathcal{A}$ and its adjoint $\mathcal{A}^{\ast}$\cite{Arridge_iter,
Haltm_iter, Scherzer_iter}, and the reconstruction times of $f(x)$ are largely governed by the computational effort spent on evaluating these operators. This computational bottleneck is exacerbated when novel model-based deep learning techniques are considered, which often require computations of $\mathcal{A}$ and
$\mathcal{A}^{\ast}$ not only in the forward pass, but also when computing gradients for neural network parameters. The total number of such evaluations during training can easily be on the order of tens of thousands or even millions
\cite{Adler2017,hauptmann2018model,arridge2019solving,hammernik2018learning}.
Hence, there is a clear need in fast algorithms for computing these operators \cite{hauptmann2018approximate,lunz2021learned}.

In this paper, we consider the practically important case of a two-dimensional measurement scheme involving circular acquisition geometry. Such a scheme is used,
for example, in the MSOT inVision scanner manufactured by iThera Medical, see
Figure 1.

\begin{figure}[t!]
\begin{center}
\includegraphics[scale = 1]{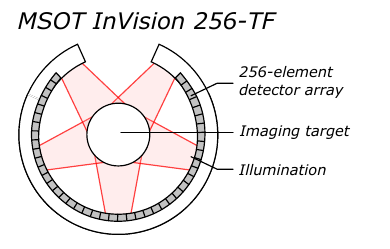}
\end{center}
\par
\vspace{-6mm}\caption{Illustration of photoacoustic system MSOT inVision by iThera Medical with the data acquired
by a ring of transducers.}
\label{F:existing}
\end{figure}

For this acquisition geometry, we develop asymptotically fast algorithms for evaluating all three operators
$\mathcal{A}$, $\mathcal{A}^{\ast}$, and $\mathcal{A}^{-1}$, and we implement
them using the Torch Python package that allows parallelized execution on a CPU
or a GPU (if present). Assuming that the dimension of the image reconstruction
grid is $(n\times n)$, there are $\mathcal{O}(n)$ transducer positions and that
$\mathcal{O}(n)$ samples are collected by each transducer, the present
algorithms run in $\mathcal{O}(n^{2}\log n)$ floating point operations (flops).
We note that asymptotically fast algorithms for the  evaluation of $\mathcal{A}$ and $\mathcal{A}^{\ast}$ in circular geometry have not previously been presented in the literature. The novel theoretically exact formulas  underlying our algorithms do not require explicit evaluation of the Hankel functions, which eliminates a possible source of instabilities, as explained in Section \ref{S:inv_adj_ops}.
We test our methods and demonstrate their performance in several numerical simulations, including
the use of the inverse operator, solution of a variational problem with and without a total variation regularizer, as well as training and application the learned primal dual method \cite{Adler2018}.

This paper is organized as follows. The next section contains a precise formulation of the problem in question. Section~\ref{S:algorithms} is devoted to derivation of the fast algorithms we propose and to the outline of their numerical implementation. The results of numerical simulations confirming the efficiency of the proposed methods are presented in Section~\ref{sec:numericalEval}. The next section contains conclusions, followed by an Appendix exhibiting derivations that may be omitted in the first reading.

\section{Formulation of the forward and inverse problem}
The\textbf{ inverse source problem} of PAT/TAT consists of reconstructing
initial pressure $f(x)$ from the measurements $g(t,y)$. This inverse problem
also arises in magnetoacoustoelectric tomography\cite{Ku-MAET}, acoustoelectric\cite{KK-AET-11} and
ultrasound-modulated optical\cite{KK-AET-10} tomography, and other hybrid modalities.
Various explicit inversion formulas and efficient
algorithms for solving this problem are known. They yield either
theoretically exact or microlocally accurate recovery of $f$ from the ideal
data $g$. However, these techniques do not take into account multiple data
deterioration factors affecting the real data. In particular, the speed of
sound may be varying across the object and not exactly known; the sound waves
are attenuated on its way to transducers; the transducers' sensitivity is
band-limited resulting in missing both high and low frequencies, etc.

We distinguish the ideal data $g^{\mathrm{ideal}}\in Y$ produced by applying
the wave operator $\mathcal{A}\colon X\to Y$ to the initial pressure $f\in X$ supported in
$\Omega_{0}$:
\[
\mathcal{A}:f\mapsto g^{\mathrm{ideal}},
\]
from real data $g$, where spaces $X$ and $Y$ specific to this paper will be defined later in the text. To make consideration more general, we can incorporate
into our equations the simplest model of acoustic attenuation by multiplying the
ideal data by the factor $e^{-\gamma t}$ with some $\gamma$ that should be found
experimentally. We will also assume that transducers have a frequently dependent
response $\eta(\xi)$, so that the real data $g$ are related to the ideal data
$g^{\mathrm{ideal}}$ as follows:
\begin{equation}
g(t,y)=[\mathcal{D}\left(  e^{-\gamma t}g^{\mathrm{ideal}}(t,y)\right)
](t,y)+\chi(t,y),\quad(t,y)\in\mathbb{R}^{+}\times\partial\Omega,
\label{E:real_meas}
\end{equation}
where $\chi(t,y)$ is the realization of a random variable that models measurement noise, and where $\mathcal{D}$ is a linear filter in the frequency
domain defined through the 1D Fourier transform $\mathcal{F}_{1D}$ in $t,$ and
its inverse $\mathcal{F}_{1D}^{-1}$:
\[
\lbrack\mathcal{D}h](t)=[\mathcal{F}_{1D}^{-1}\left(  \eta(\xi)[\mathcal{F}
_{1D}h](\xi)\right)  ](t).
\]
Formula (\ref{E:real_meas}) (with $\chi(t,y)\equiv0$) defines an operator
\[
\mathcal{B}:g^{\mathrm{ideal}}\mapsto g,
\]
and the relation between $f$ and $g$ is given by the following composition:
\[
g(t,y)=[\mathcal{BA}f](t,y)+\chi(t,y),\quad(t,y)\in\mathbb{R}^{+}
\times\partial\Omega.
\]

Note that, depending on the properties of $\eta(\xi)$, the mutual
location of $\Gamma$, and the support of $f$, one or both operators $\mathcal{A}$
and $\mathcal{B}$ may be not invertible or not stably invertible,
especially in the presence of noise $\chi(t,y)$. As a result, in practical
applications of TAT/PAT dealing with real data $g(t,y)$, the beauty and
computational efficiency of explicit reconstruction techniques is frequently
sacrificed in favor of non-linear optimization algorithms and/or deep learning
techniques that produce images of superior quality.
We note that evaluation of $\mathcal{B}$ and $\mathcal{B}^{\ast}$ is
relatively simple and we will omit these operators from further discussion.

Let us now discuss established image reconstruction approaches that are used when the reconstruction problem becomes more ill-posed, i.e., under limited-view and/or high-noise scenarios.
First, classical variational methods seek to find an
approximation $\tilde{f}(x)$ as a minimizer of a suitable cost functional, such as
\begin{equation}
\tilde{f}=\argmin_{f\geq 0} \|\mathcal{A}f-g\|_{2}^{2}+\alpha \mathcal{R}(f),
\label{E:minim}
\end{equation}
where the first term measures data consistency, $\mathcal{R}$ is a regularizing term, and $\alpha$ balances the two terms. There is a wide variety of different
choices of $\mathcal{R}$ enforcing regularity $\tilde{f}(x)$, such as smoothness and sparsity with respect to a suitable transform or closeness to a known
prior. Minimization of the functional (\ref{E:minim}), as a rule, is done using iterative optimization techniques, which usually require the computation of a descent direction in each iteration. The simplest of such methods uses gradient information and necessitates computation of the gradient of the data fidelity given by
\begin{equation}
\nabla \|\mathcal{A}f-g\|_{2}^{2} = 2\mathcal{A}^{\ast}(\mathcal{A}f-g).
\label{E:operators}
\end{equation}
We note that the splitting techniques and primal-dual methods that are commonly used to minimize \eqref{E:minim} also involve the computation of $\mathcal{A}$ and $\mathcal{A}^*$ in each iteration.
For all such methods, the number of iterations
needed to achieve convergence is on the order of hundreds or even thousands. This amounts to a significant computational effort, mainly spent on the evaluation of
$\mathcal{A}$ and its adjoint~$\mathcal{A}^{\ast}$.

Recently, several successful applications of deep learning techniques to the
inverse problem of PAT/TAT have been reported \cite{grohl2021deep,hauptmann2020deep,yang2021review}, training neural networks on realistic (or better yet, real) images
allows one to promote realistic image patterns that are not easily expressed in terms of smoothness, total variation, and sparsity.
Among these techniques are so-called model-based learned iterative methods
\cite{hauptmann2018model,hauptmann2018approximate},
which formulate the iteration process in the following form:
\[
f^{(k+1)}=\Lambda_{\theta_{k}}(f^{(k)},\mathcal{A}^{\ast}(\mathcal{A}f^{(k)}-g)),
\]
where $\Lambda_{\theta_{k}}$ is a convolutional neural network with a set
$\theta_{k}$ of learned parameters, trained to improve the image generated on iteration
number $k$. This technique was used successfully in \cite{hauptmann2018model,hauptmann2018approximate,hsu2021comparing} to remove artifacts caused by an insufficient field of view; it also showed promise to correct, additionally, for many other image deterioration factors.
The above scheme was trained by greedy training, a computational trick that trains each iterate separately and decouples the evaluation of $\mathcal{A}$ and $\mathcal{A}^\ast$ from training of the network parameters. The whole training took 5 days, where the bulk of the computational effort was spent, again, on numerical evaluation of
the operators $\mathcal{A}$ and $\mathcal{A}^{\ast}$.

In the case of linear and planar acquisition geometries, a faster training \cite{hauptmann2018approximate}, as well as the end-to-end training of all iterates simultaneously \cite{hauptmann2023model} were achieved by using efficient FFT-based algorithms.
Thus in commonly used circular and spherical geometries, to apply modern optimization techniques
and/or investigate advantages of deep learning-assisted TAT and PAT  one needs
the fastest possible algorithms for the evaluation of $\mathcal{A}$ and $\mathcal{A}^{\ast}$.

\subsection{Known methods}
The properties of the three operators of interest, $\mathcal{A}\colon X\to Y$, $\mathcal{A}^\ast\colon Y\to X$, and $\mathcal{A}^{-1}\colon Y\to X$, depend on the geometry of the measuring scheme.
Current commercially available PAT scanners either use a spherical data
acquisition surface (such as LOIS 3D by Tomowave) or utilize a ring of
detectors moving along the object of interest (MSOTinVision by iThera
Medical), as shown in Figure \ref{F:existing}. In the present paper, we
concentrate on the 2D circular acquisition geometry, leaving the 3D spherical
acquisition for future work. We consider the acquisition geometry where
$f$ is supported within a disk $\Omega$ with boundary $S$ and the pressure
is measured on a subset of its boundary $\Gamma\subset S$. Without loss of
generality, one can assume that the circle is of radius 1 and that the speed of
sound also equals 1. Under these assumptions, the forward operator
$\mathcal{A}$ can be expressed as the convolution of $f$ with the fundamental
solution of the free-space wave equation $G(t,x)$:
\begin{align}
g(t,z)  &  =[\mathcal{A}f](t,z)=\frac{\partial}{\partial t}\int\limits_{\Omega
_{0}}f(x)G(t,x-z)dx,\qquad(t,x)\in Q\equiv(0,T)\times\Gamma
,\label{E:green_conv}\\
G(t,x)  &  =\frac{1}{2\pi\sqrt{t^{2}-|x|^{2}}}\text{ for }t>|x|,\text{ and
}0\text{ otherwise,}\nonumber
\end{align}
where $Q$ is the time-space cylinder. Further, by defining the $L^2$ inner products
$\langle\cdot,\cdot\rangle_{\Omega_{0}}$ and $\langle\cdot,\cdot\rangle_{Q}$ on $\Omega_0$ and $Q$, respectively, as
\begin{equation}\label{eqn:innerProducts}
\langle u,v\rangle_{\Omega_0}=\int\limits_{\Omega_{0}}u(x)v(x)\,dx,\qquad \langle h,k\rangle_{Q}
=\int\limits_{0}^{T}\int\limits_{\Gamma}h(t,z)k(t,z)\,dz\,dt,
\end{equation}
one obtains the expression for the adjoint operator $\mathcal{A}^{\ast}$:
\begin{equation}
\lbrack\mathcal{A}^{\ast}g](x)=\int\limits_{0}^{T}\int\limits_{\Gamma
}g(t,z)\frac{\partial}{\partial t}G(t,x-z)\,dz\,dt,\qquad x\in\Omega_{0}.
\label{E:adjoint}
\end{equation}
The above formula is equivalent to the previous definitions of the adjoint operator
(e.g. \cite{Haltm_iter,Arridge_iter,Scherzer_iter}) if the speed of sound is
assumed constant. Additionally, the inner products in \eqref{eqn:innerProducts} define the function spaces for the reconstructions $f\in X:=L^2(\Omega_0)$ as well as for the measurements $g\in Y:= L^2(Q)$.

Finally, it is known\cite{Haltm-universal-SIAM} that the following formula defines a left inverse
operator for $\mathcal{A}$:
\begin{equation}
f(x)=[\mathcal{A}^{-1}g](x)\equiv2\int\limits_{0}^{\infty}\int\limits_{S
}g(t,z)\frac{\partial}{\partial n(z)}G(t,x-z)\,dz\,dt, \label{E:inverse}
\end{equation}
where $n(z)$ is the exterior normal to $S$ at the point $z.$
\subsubsection{Forward operator\label{S:known_forward}}

Early works on PAT and TAT\cite{hkn,BurgPhysRev} modeled the forward operator
by solving the Cauchy problem (\ref{E:wave-eq}), (\ref{E:Cauchy}) using second-order finite differences. Such an algorithm requires $\mathcal{O}(n^{2})$
flops per time step, and, assuming that $\mathcal{O}(n)$ time steps need
to be computed, the total computational cost of the method is $\mathcal{O}
(n^{3})$ flops. Such an operation count is too high for the purposes of this
paper. In addition, the accuracy of the second-order finite differences is
quite low if the function $f(x)$ representing the initial pressure undergoes
sharp changes (e.g., see Section \ref{S:complete} and Figure 3(d) for the example of artifacts resulting from the use of the finite differences).

Frequently, the computation of the forward operator $\mathcal{A}$ in PAT and
TAT is based on Fourier transform methods, accelerated using
the Fast Fourier Transform algorithm (FFT). Let us define the Fourier
transform $\mathcal{F}_{2D}$
and its inverse $\mathcal{F}_{2D}^{-1}$ by the following formulas:
\[
\hat{h}(\xi)=[\mathcal{F}_{2D}h](\xi)\equiv\frac{1}{2\pi}\int
\limits_{\mathbb{R}^{2}}h(x)\exp(-i\xi\cdot x)\,dx,\quad h(x)=[\mathcal{F}
_{2D}^{-1}h](x)\equiv\frac{1}{2\pi}\int\limits_{\mathbb{R}^{2}}\hat{h}
(\xi)\exp(i\xi\cdot x)\,d\xi,
\]
where function $h(x)$ is an element of the Schwartz space on $\mathbb{R}^{2}$\cite{NattBook} .
Then the solution of the forward problem (\ref{E:wave-eq}), (\ref{E:Cauchy})
can be written as follows (e.g. Chap. 3 in \cite{rauchPDE}):
\begin{align}
p(t,x)  &  =\left[  \mathcal{F}_{2D}^{-1}\left(  \hat{f}(\xi)\cos
(|\xi|t)\right)  \right]  (t,x),\qquad(t,x)\in(0,\infty)\times\mathbb{R}
^{2},\label{E:kwave}\\
\hat{f}(\xi)  &  =[\mathcal{F}_{2D}f](\xi),\nonumber
\end{align}
where we assumed that $f(x)\ $is extended by 0 to all of $\mathbb{R}^{2}$.
This leads to a relatively simple algorithm. First, we precompute $\hat
{f}(\xi)$ and choose discretization for time $t$. Then for each value of the
variable $t$, the inverse Fourier transform (\ref{E:kwave}) is calculated to
produce the values of $p(t,x)$ on a Cartesian grid in $x$. Finally, the values
of $g(t,z)$ at the transducer points discretizing the unit circle are computed
by interpolation from $p(t,x)$. Since $f(x)$ is finitely supported, the
computation of both $\hat{f}(\xi)$ and $p(t,x)$ is spectrally accurate, and
the total precision of the algorithm depends on the order of the interpolation.
This can be viewed as a simplified version of the popular \emph{k-Wave}
algorithm \cite{kwave}. (The latter algorithm is more advanced in that it can
handle an inhomogeneous speed of sound, but in the case of the constant speed
it reduces to what is described here). This approach is frequently used by
researchers (e.g. \cite{Haltm_iter,Arridge_iter,Scherzer_iter}).

The unaccelerated Fourier algorithm based on equation (\ref{E:kwave}) requires
$\mathcal{O}(n^{2}\log n)$ flops to produce $p(t,x)$ at each step in$~t$.
Assuming that $\mathcal{O}(n)$ time steps need to be computed, the total
complexity is $\mathcal{O}(n^{3}\log n)$ flops. This is slower than the
accelerated $\mathcal{O}(n^{2}\log n)$ flops algorithm we present further in
this paper. However, we will use this unaccelerated Fourier algorithm to
verify the accuracy of the fast method we propose.

\subsubsection{Inverse and adjoint operators}\label{S:inv_adj_ops}

Known PAT/TAT image reconstruction algorithms, based on
forward/adjoint iterations (\cite{Haltm_iter,Arridge_iter,Scherzer_iter}), use
a modification of the \emph{k-Wave} algorithm \cite{kwave} discussed above to
compute the result of the application of the adjoint operator $\mathcal{A}^{\ast}$
to the data supported on the time-space cylinder $Q$. Without going into
details, the operation count for such an approach is $\mathcal{O}(n^{3}\log
n)$ flops. Our goal is to develop methods for evaluating
$\mathcal{A}$ and $\mathcal{A}^{\ast}$ in $\mathcal{O}(n^{2}\log n)$ flops.

Let us start, however, with the inverse operator $\mathcal{A}^{-1}$. Applications of $\mathcal{A}^{-1}$ to data $g$ given on
$Q$ have received significant attention from the researchers, because, in the
case of complete data $g$ (for example, when $\Gamma$ covers the whole circle
$S$ and $T$ is infinite), computation of $\mathcal{A}^{-1}g$ recovers the
sought initial pressure $f.$ We note that, in fact, the left inverse operator
$\mathcal{A}^{-1}$ is not unique; in the case of ideal data there exist many
non-equivalent operators that recover $f$ from $\mathcal{A}f$ (see, e.g.
\cite{Nguyen_fam}). We will not try to overview the vast literature on the
inverses and will only concentrate on known fast algorithms.

For the simpler case where the acquisition surface $\Gamma$ is a line or a
plane, Fourier-based algorithms with complexity $\mathcal{O}(n^{d}\log n)$
(where $d$ is the dimension of the space) have been developed
\cite{Kostli_FFT,Haltm_FFT}. However, in this paper, we are interested in a
circular acquisition geometry. For the latter scheme, a fast image
reconstruction algorithm (that can be viewed as a scheme to compute a particular instance of $\mathcal{A}^{-1}$) was developed in \cite{Kun-cyl} and applied to measurement data in \cite{DigitalTwin2025}.

Let us review the basics of the latter method. By
extending $g(t,z)$ by zero in $t$ to a function defined on $\mathbb{R}\times
S$ one can Fourier transform $p$ in time; we will denote the result by
$\hat{g}(\lambda,z)$:
\begin{equation}
\hat{g}(\lambda,z)\equiv\int\limits_{\mathbb{R}}g(t,z)e^{it\lambda}dt,\qquad
z\in S. \label{Phat2D}
\end{equation}
Further, let us switch to polar coordinates in $x$:
\[
x=x(r,\theta)=r(\cos\theta,\sin\theta),r=|x|,
\]
and expand $\hat{g}(t,z(\varphi)),$ $z=(\cos\varphi,\sin\varphi)$, and
$f(r\hat{x}(\theta))$ into the Fourier series in $\varphi$ and $\theta$:
\begin{equation}
\hat{g}(\lambda,z(\varphi))=\sum\limits_{k=-\infty}^{\infty}\hat{g}
_{k}(\lambda)e^{ik\varphi},\qquad f(r\hat{x}(\theta))=\sum\limits_{m=-\infty
}^{\infty}f_{m}(r)e^{im\theta}.\nonumber
\end{equation}
where $\hat{g}_{k}(\lambda)$ and $f_{m}(r)$ are the corresponding Fourier series coefficients, $k\in\mathbb{Z}$. Then \cite{Kun-cyl} these coefficients
are related through the following formula:
\begin{equation}
\hat{g}_{k}(\lambda)=\frac{\pi}{2}\lambda H_{|k|}^{(1)}(\lambda)\int
\limits_{0}^{\infty}f_{|k|}(r)J_{|k|}(\lambda r)rdr, \label{E:Hankels}
\end{equation}
where $J_{k}(\cdot)$ are $H_{k}^{(1)}(\cdot)$ are the Bessel functions and
Hankel functions of the first kind, of order $k$, and where the integrals are
the Hankel transforms of order $|k|$. In \cite{Kun-cyl}, to
reconstruct $f$ from $g$, one computes $\hat{g}_{k}(\lambda)$ and finds the
values of the integrals in equation (\ref{E:Hankels}) by division:
\begin{equation}
\int\limits_{0}^{\infty}f_{|k|}(r)J_{|k|}(\lambda r)rdr=\frac{2\hat{g}
_{k}(\lambda)}{\pi\lambda H_{|k|}^{(1)}(\lambda)},\qquad k\in\mathbb{Z}
,\qquad\lambda\in\mathbb{R}\backslash\{0\}. \label{E:Han_inverse}
\end{equation}
The Hankel functions in the denominator of the above formula do not have zeros
for real values of $\lambda$, so the division is well defined. After the
values of Hankel transforms in (\ref{E:Han_inverse}) are computed, $f$ is
reconstructed using the well-known connection between the Hankel transforms
and the Fourier series of $\hat{f}\equiv\mathcal{F}_{2D}f$ considered in polar
coordinates, see \cite{Kun-cyl} for details.

One may wonder whether a fast algorithm for computing $\mathcal{A}f$ can be
developed by performing the above steps in reverse. Theoretically, this seems
to be possible. Direct computation of Hankel transforms would require
$\mathcal{O}(n^{3})$ flops, which is too expensive for our purposes. Instead,
one could compute these transforms of $f$ indirectly, through the connection
with the Fourier transform $\hat{f}$ of $f$. However, a problem will arise
when formula (\ref{E:Hankels}) is used to compute coefficients $\hat{g}
_{k}(\lambda)$. Namely, for a fixed $\lambda$ and large order $k$, Hankel
functions exhibit very fast growth (see formula 10.19.2 in
\cite{NIST_handbook}) that starts roughly at $|k|\geq|\lambda|$:
\[
|H_{|k|}^{(1)}(\lambda)|\thicksim\sqrt{\frac{2}{\pi|k|}}\left(  \frac
{2|k|}{e|\lambda|}\right)  ^{k}\text{ as }|k|\rightarrow\infty.
\]
Ideally, this growth is compensated for by the rapid decay of the Bessel functions
(\ref{E:Hankels}) at the limit $|k|\rightarrow\infty$. However, if the Hankel
transforms are computed indirectly, there is no reason to expect a fast decay
of absolute error in computed Hankel transforms with $|k|\rightarrow\infty$.
Therefore, such an algorithm for computing $\mathcal{A}f$ would be unstable.

Similarly, exploiting equation (\ref{E:Hankels}) as a starting point for an
algorithm for computing $\mathcal{A}^{\ast}$ also leads to multiplication of
approximately computed quantities by Hankel functions. Thus, below we find a
different strategy for a fast computation of $\mathcal{A}$, $\mathcal{A}
^{\ast}$ and $\mathcal{A}^{-1}\mathcal{\ (}$defined by (\ref{E:inverse})) that
does not involve Hankel functions at all.

\section{Fast algorithms for computing $\mathcal{A}$,\ $\mathcal{A}^{\ast}$
and $\mathcal{A}^{-1}$}\label{S:algorithms}
In the present section we derive the algorithms for evaluation of the forward operator $\mathcal{A}\colon X\to Y$ and its adjoint $\mathcal{A}^\ast\colon Y\to X$, and describe their implementation. We also present an implementation of the fast Hankel-free version of the inverse $\mathcal{A}^{-1}\colon Y\to X$, but the derivation of the underlying formulas is relegated to the Appendix.

Our codes are made available on Github at \url{https://github.com/leonidak/torch-TAT}, together with computational examples demonstrating the use of the proposed fast algorithms.

\subsection{Forward operator\label{S:forward}}

Starting with formula (\ref{E:kwave}), for any point $\hat{z}(\theta
)=(\cos\theta,\sin\theta)$ lying on the unit circle $\mathbb{S}$ we obtain:
\begin{equation}
g(t,\hat{z}(\theta))=\frac{1}{2\pi}\int\limits_{\mathbb{R}^{2}}\hat{f}
(\xi)\cos(|\xi|t)\exp(i\xi\cdot z(\theta))\,d\xi. \label{E:gseries}
\end{equation}
Let us expand this function in the Fouirer series in $\theta$:
\begin{equation}
g(t,\hat{z}(\theta))=\sum_{k=-\infty}^{\infty}g_{k}(t)e^{ik\theta},\qquad
g_{k}(t)=\frac{1}{2\pi}\int\limits_{0}^{2\pi}g(t,\hat{z}(\theta))e^{-ik\theta
}\,d\theta. \label{E:FourSerForg}
\end{equation}
Substitute (\ref{E:gseries}) into (\ref{E:FourSerForg}) and switch to polar
coordinates $\xi=|\lambda|\hat{\xi}$, $\hat{\xi}=(\cos(\varphi,\sin\varphi))$:
\begin{align}
g_{k}(t)  &  =\frac{1}{\left(  2\pi\right)  ^{2}}\int\limits_{0}^{2\pi}\left(
\int\limits_{\mathbb{R}^{2}}\hat{f}(\xi)\cos(|\xi|t)\exp(i\xi\cdot\hat
{z}(\theta))\,d\xi\right)  e^{-ik\theta}\,d\theta\nonumber\\
&  =\frac{1}{\left(  2\pi\right)  ^{2}}\int\limits_{\mathbb{R}^{2}}\hat{f}
(\xi)\cos(|\xi|t)\left(  \int\limits_{0}^{2\pi}\exp(i\xi\cdot\hat{z}
(\theta))\,e^{-ik\theta}\,d\theta\right)  \,d\xi\nonumber\\
&  =\frac{1}{\left(  2\pi\right)  ^{2}}\int\limits_{0}^{\infty}\left[
\int\limits_{0}^{2\pi}\hat{f}(\lambda\hat{\xi}(\varphi))\cos(\lambda t)\left(
\int\limits_{0}^{2\pi}\exp(i\lambda\hat{\xi}(\varphi)\cdot\hat{z}
(\theta))\,e^{-ik\theta}\,d\theta\right)  \,d\varphi\right]  \,\lambda
d\lambda. \label{E:long_integral}
\end{align}
We will need to simplify the inner integral in parentheses in
(\ref{E:long_integral}):
\begin{equation}
\int\limits_{0}^{2\pi}\exp(i\lambda\hat{\xi}(\varphi)\cdot\hat{z}
(\theta))\,e^{-ik\theta}\,d\theta=\int\limits_{0}^{2\pi}\exp(i\lambda
\cos(\theta-\varphi))e^{-ik\theta}\,d\theta=e^{-ik\varphi}\int\limits_{0}
^{2\pi}e^{i\lambda\cos\theta}e^{-ik\theta}\,d\theta. \label{E:eq1}
\end{equation}
This can be done by substituting into (\ref{E:eq1}) the Jacobi-Anger expansion
(see, e.g. \cite{Colton}):
\begin{equation}
e^{i\lambda\cos\theta}=J_{0}(\lambda)+2\sum\limits_{n=1}^{\infty}i^{n}
J_{n}(\lambda)\cos(n\theta)=\sum\limits_{n=-\infty}^{\infty}i^{|n|}
J_{|n|}(\lambda)e^{in\theta}, \label{E:Anger}
\end{equation}
which results in the following formula:
\[
\int\limits_{0}^{2\pi}\exp(i\lambda\hat{\xi}(\varphi)\cdot\hat{z}
(\theta))\,e^{-ik\theta}\,d\theta=2\pi e^{-ik\varphi}i^{|k|}J_{|k|}(\lambda).
\]
Combining the latter formula with (\ref{E:long_integral}) we obtain the
expression for $g_{k}(t)$:
\[
g_{k}(t)=i^{|k|}\int\limits_{0}^{\infty}\left[  \frac{1}{2\pi}\int
\limits_{0}^{2\pi}\hat{f}(\lambda\hat{\xi}(\varphi))e^{-ik\varphi}
\,d\varphi\right]  \,J_{|k|}(\lambda)\cos(\lambda t)\,\lambda d\lambda.
\]
The expression in brackets defines the Fourier coefficients $\hat{f}
_{k}(\lambda)$of $\hat{f}(\lambda\hat{\xi}(\varphi))$ with respect to
$\varphi$:
\begin{equation}
\hat{f}_{k}(\lambda)=\frac{1}{2\pi}\int\limits_{0}^{2\pi}\hat{f}(\lambda
,\hat{\xi}(\varphi))e^{-ik\varphi}\,d\varphi, \label{E:eq3}
\end{equation}
so that the Fourier coefficients $g_{k}$ of $g$ are expressed through the
coefficients $\hat{f}_{k}(\lambda)$:
\begin{equation}
g_{k}(t)=i^{|k|}\int\limits_{0}^{\infty}\left[  \lambda\hat{f}_{k}
(\lambda)\,J_{|k|}(\lambda)\right]  \cos(\lambda t)\,d\lambda. \label{E:eq2}
\end{equation}
Recall that the cosine Fourier transform $[\mathcal{F}_{\cos}h](s)$ of
function $h(y)$ is defined as follows:
\[
\lbrack\mathcal{F}_{\cos}h](s)=\sqrt{\frac{2}{\pi}}\int\limits_{0}^{\infty
}h(y)\cos(sy)\,dy.
\]
Thus, equation (\ref{E:eq2}) can be interpreted as the cosine Fourier
transform of $\lambda\hat{f}_{k}(\lambda)\,J_{|k|}(\lambda)$:
\begin{equation}
g_{k}(t)=i^{|k|}\sqrt{\frac{\pi}{2}}\left[  \mathcal{F}_{\cos}(\lambda\hat
{f}_{k}(\lambda)\,J_{|k|}(\lambda))\right]  (t). \label{E:inverse_cos_tr}
\end{equation}
We observe that once $\hat{f}(\xi)$ is computed, Fourier coefficients
$\hat{f}_{k}(\lambda)$ can be computed using (\ref{E:eq3}) and coefficients
$g_{k}(t)$ can be obtained by evaluating cosine Fourier transforms
(\ref{E:inverse_cos_tr}). The sought function $g(t,z)$ is then reconstructed
by summing the Fourier series (\ref{E:FourSerForg}).

Our fast algorithm for computing $g=\mathcal{A}f$ results by replacing the
Fourier transforms and Fourier series by their discrete counterparts.
Computation of the Fourier coefficients and summation of the Fourier series of
periodic functions using the FFT is straightforward. On the other hand,
replacement of the Fourier transforms by the FFT's requires some discussion.
Replacing in (\ref{E:kwave}) the true Fourier transform $\mathcal{F}_{2D}$
computed over $\mathbb{R}^{2},$ by the FFT sampling the square $\mathfrak{S}
\equiv\lbrack-L,L]\times\lbrack-L,L]$ results in the solution of the wave equation that is periodic in $\mathfrak{S}$. This means that so computed approximation
to $g=\mathcal{A}f$ will become incorrect when the support of the solution,
initially contained inside $S,$ will reach the boundaries of $\mathfrak{S}$,
reflects back and reaches $S.$ Since the speed of sound in our model is equal
to $1,$ this will happen at $T_{\mathrm{refl}}=2(L-1)$. Therefore, the
parameter $L$ of $\mathfrak{S}$ should be chosen so that $T_{\mathrm{refl}
}>T,$ or
\[
L>\frac{T}{2}+1.
\]
(In fact, the above formula yields a very minimal value for $L$. As explained below, parameter $L$ may require even further increase).

Similarly to the 2D FFT's, replacing the cosine transform $\mathcal{F}_{\cos}$
in (\ref{E:inverse_cos_tr}) by its discrete version will produce a periodic
approximation to all coefficients $g_{k}(t)$. Since the wave equation in 2D
does not obey the Huygens principle, the solution will have relatively small tails, slowly decreasing in time. In a periodic approximation, these tails
will wrap around, producing a noticeable error. In order to decrease this
effect, we increase the model time $T$ making $T_{\mathrm{new}}=\max(2T,6).$
Moreover, for the lowest circular harmonics $g_{k}(t)$ with $k=-1,0,1$ we do
not use the discrete cosine FFT and compute integrals in
(\ref{E:inverse_cos_tr}) using a quadrature rule with discretization points
clustering toward $\lambda=0.$

Finally, interpolation of the values of $\hat{f}(\xi)$ computed by a 2D\ FFT
on a Cartesian grid in $\xi,$ to the polar grid also requires a discussion. It
is well known that the low order interpolation in the Fourier domain results
in inaccurate images (see, e.g., Ch. V.2 in \cite{NattBook}). Fortunately, the
Fourier transform of a compactly supported function is a band-limited function,
implying that higher-order methods should work very well. Perhaps the most
accurate interpolation techniques are those based on the nonuniform FFT (NUFFT), (e.g., \cite{Greengard}).
However, while formally such a step would be within the desired $\mathcal{O}
(n^{2}\log n)$ flops count, the constant factor hidden in the $\mathcal{O}
(...)$ notation is quite large for various versions of NUFFTs. A faster method, well-suitable for a single
thread computation is bi-cubic interpolation. When experimenting with our code,
we obtained sufficient accuracy with the help of the bi-cubic interpolation
routines present in the SciPy package in Python. (By sufficient accuracy we
understand relative $L^{\infty}$ errors of order of a half-of-a-percent or
less). Unfortunately, such an interpolation is not local and, as a result, it
is difficult to parallelize. For the numerical simulations presented in
Section \ref{sec:numericalEval}, we combined refining the discretization of the Cartesian
frequency grid (obtained by doubling the size of $L$ to the value
$L\thickapprox T_{new}+2$) with consecutive bi-linear interpolation. On one
hand, this produced an accuracy comparable with that of bi-cubic
interpolation. On the other hand, bi-linear interpolation is easily
vectorizable and parallelizable. This allowed us to implement the algorithm
using the Python Torch package that provides automatic parallelization and
execution on a GPU (if present).
This results in the Algorithm \ref{alg:forward}.
\begin{algorithm}[h!]
\caption{Computing $g = \mathcal{A}f$}
\label{alg:forward}
\begin{algorithmic}[1]

\item Extend $f(x)$ by zero to a larger square domain $\mathfrak{S;}$

\item Using the 2D FFT compute an approximation to $\hat{f}(\xi)$ on a
Cartesian grid in $\xi$;

\item Use bi-linear interpolation to obtain values of $\hat{f}(\lambda\hat
{\xi}(\varphi))$ on the polar grid;

\item For each value $\lambda$ in the polar grid in $\xi$, compute Fourier
coefficients $\hat{f}_{k}(\lambda)$ using 1D FFT;

\item Compute Fourier coefficients $g_{k}(t)$ using the discrete Fourier
cosine transform, equation (\ref{E:inverse_cos_tr}), for each time step in $t$;

\item For each time step $t$, compute $g(t,\hat{z}(\theta))$ by the FFT in
angle $\theta$, see equation (\ref{E:FourSerForg});

\item Restrict $g(t,\hat{z}(\theta))$ to $\Gamma_{.}$
\end{algorithmic}
\end{algorithm}

The most time-consuming steps of the algorithm are the computation of 2D FFT
which requires $\mathcal{O}(n^{2}\log n)$ flops, and $\mathcal{O}(n)$
evaluations of 1D FFT's which has the similar operation count. So, the whole
algorithm needs $\mathcal{O}(n^{2}\log n)$ flops.

Since this algorithm is designed mainly for use in iterations, additional time
saving is achieved by pre-computing and storing the values of the Bessel
functions $J_{|k|}(\lambda)$ for the values of $k$ from $0$ to half the number of discretization points in $\theta$, and for all values of
$\lambda=|\xi|$ in the selected polar grid in $\xi$. Further details of the implementation can be found in Section \ref{sec:numericalEval}.

\subsection{Adjoint operator}

The integral in equation (\ref{E:adjoint}) describes wave propagation in
the whole of $\mathbb{R}^{2}$ excited by a source supported on $Q,$ backwards in time from $t=T$ to $0$. We will denote the whole acoustic field at $t=0$ by
$u(x),$ so that the adjoint operator can be written as follows:
\begin{equation}
\lbrack\mathcal{A}^{\ast}g](x)=\left[  \Pi_{\Omega_{0}}u\right]  (x),\qquad
u(x)\equiv\int\limits_{0}^{T}\int\limits_{S}g(t,z)G_{t}^{\prime}
(t,x-z)\,dz\,dt,\quad x\in\mathbb{R}^{2}, \label{E:new_adjoint}
\end{equation}
where $\Pi_{\Omega_{0}}$ is a projection operator that restricts a function
from $\mathbb{R}^{2}$ to $\Omega_{0}$, and where (with abuse of notation) we
extended $g(t,z)$ by zero from $\Gamma$ to $S$. For simplicity, we will assume
that the distance from $\Omega_{0}$ to $\partial\Omega$ is nonzero.

The free-space fundamental solution $G(t,x)$ can be expressed through its
Fourier transform \cite{Eskin}:
\begin{equation}
G(t,x)=\left[  \mathcal{F}_{2D}^{-1}\left(  \frac{\sin|\xi|t}{|\xi|}\right)
\right]  (t,x),\text{\qquad}G_{t}^{\prime}(t,x)=\left[  \mathcal{F}_{2D}
^{-1}\left(  \cos|\xi|t\right)  \right]  (t,x),\text{ for }t>0.
\label{E:Fourier_Green}
\end{equation}
By substituting the formula for $G_{t}^{\prime}(t,x)$ into equation
(\ref{E:new_adjoint}) one obtains:
\begin{align*}
u(x)  &  =\int\limits_{0}^{T}\int\limits_{0}^{2\pi}g(t,\hat{z}(\theta))\left(
\frac{1}{2\pi}\int\limits_{\mathbb{R}^{2}}\cos(|\xi|t)\,e^{i\xi\cdot
(x-z)}\,d\xi\right)  \,d\theta\,dt\\
&  =\frac{1}{2\pi}\int\limits_{\mathbb{R}^{2}}\left[  \int\limits_{0}
^{T}\left(  \int\limits_{0}^{2\pi}g(t,\hat{z}(\theta))\cos(|\xi|t)\,e^{-i\xi
\cdot z}\,d\theta\right)  \,dt\right]  e^{i\xi\cdot x}\,d\xi=[\mathcal{F}
_{2D}^{-1}\hat{u}](x),
\end{align*}
where
\begin{equation}
\hat{u}(\xi)=\int\limits_{0}^{T}\left(  \int\limits_{0}^{2\pi}g(t,\hat
{z}(\theta))\cos(\lambda t)\,e^{-i\xi\cdot z}\,d\theta\right)  \,dt,\qquad
\lambda=|\xi|. \label{E:FT_u}
\end{equation}
Using the Fourier series for $g(t,z(\theta))$ (equation
(\ref{E:FourSerForg})), the inner integral (in $\theta$) in (\ref{E:FT_u}) can
be transformed as follows:
\begin{align}
\int\limits_{0}^{2\pi}g(t,\hat{z}(\theta))\cos(\lambda t)\,e^{-i\xi
z}\,d\theta &  =\int\limits_{0}^{2\pi}\left[  \sum_{k=-\infty}^{\infty}
g_{k}(t)e^{ik\theta}\right]  \cos(\lambda t)\,e^{-i\xi z}\,d\theta\nonumber\\
&  =\sum_{k=-\infty}^{\infty}g_{k}(t)\cos(\lambda t)\left(  \int
\limits_{0}^{2\pi}e^{-i\xi\cdot z}e^{ik\theta}\,d\theta\right)  ,
\label{E:eq5}
\end{align}
where $\xi=\lambda(\cos\varphi,\sin\varphi)$, $\hat{z}(\theta)=(\cos
\theta,\sin\theta)$. Now, the Jacobi-Anger expansion (\ref{E:Anger}) allows us to further simplify the integral in (\ref{E:eq5}):
\begin{align*}
\int\limits_{0}^{2\pi}e^{-i\xi\cdot z}e^{ik\theta}\,d\theta &  =\int
\limits_{0}^{2\pi}e^{-i\lambda\cos(\theta-\varphi)}e^{ik\theta}\,d\theta
=e^{ik\varphi}\int\limits_{0}^{2\pi}\,e^{-i\lambda\cos\theta}e^{ik\theta
}\,d\theta.\\
&  =e^{ik\varphi}\int\limits_{0}^{2\pi}\,\left[  \sum\limits_{n=-\infty
}^{\infty}i^{|n|}J_{|n|}(-\lambda)e^{in\theta}\right]  e^{ik\theta}
\,d\theta=2\pi e^{ik\varphi}(-i)^{|k|}J_{|k|}(\lambda).
\end{align*}
Thus, for the inner integral in (\ref{E:FT_u}) we obtain the following
representation:
\[
\int\limits_{0}^{2\pi}g(t,\hat{z}(\theta))\cos(\lambda t)\,e^{-i\xi\cdot
z}\,d\theta=2\pi\sum_{k=-\infty}^{\infty}e^{ik\varphi}(-i)^{|k|}
g_{k}(t)J_{|k|}(\lambda)\cos(\lambda t),
\]
so that $\hat{u}(\xi)$ can be computed by the formulas:
\begin{align}
\hat{u}(\xi(\lambda,\varphi))  &  =\sum_{k=-\infty}^{\infty}e^{ik\varphi}
\hat{u}_{k}(\lambda),\label{E:FourSerForu}\\
\hat{u}_{k}(\lambda)  &  \equiv2\pi(-i)^{|k|}J_{|k|}(\lambda)\int
\limits_{0}^{T}g_{k}(t)\cos(\lambda t)\,dt. \label{E:Four_coef_u}
\end{align}
It is evident from the above equation that the functions $\hat{u}_{k}
(\lambda)$ we just introduced are the coefficients of expansion of $\hat
{u}(\xi)=\hat{u}(\lambda(\cos\varphi,\sin\varphi))$ in the Fourier series in
$\varphi$. Thus, we have reduced the computation of $\mathcal{A}^{\ast}g$ to
evaluations of Fourier series, Fourier cosine transforms, and one 2D inverse
Fourier transform.

As in the case of the forward operator $\mathcal{A},$ in order to facilitate
good accuracy of bilinear interpolations and accurate representation of
non-periodic functions by FFT's that are naturally periodic, we define an
extended Cartesian grid of the size $[-L,L]\times\lbrack L,L]$ with $L=1.1+T$,
and we extend the data $g$ by zeros to a larger time interval $t\in
\lbrack0,T_{\mathrm{large}}]$ with $T_{\mathrm{large}} = \max(2.1,4T)$.
This results in Algorithm \ref{alg:adjoint}.
\begin{algorithm}[h!]
\caption{Computing $u = \mathcal{A^*}g$}
\label{alg:adjoint}
\begin{algorithmic}[1]

\item Extend $g(t,x)$ by zero from $(0,T)\times\Gamma$ to
$(0,T_{\mathrm{large}})\times S$;

\item Using the 1D FFTs compute coefficients $g_{k}(t)$ (equation
(\ref{E:FourSerForg})) for each grid value of $t$;

\item Using the 1D cosine FFTs compute coefficients $\hat{u}_{k}(\lambda)$ for
each grid value of $\lambda$, equation (\ref{E:Four_coef_u});

\item Using the 1D FFTs, sum the Fourier series (\ref{E:FourSerForu}) for each
grid value of $\lambda$, thus obtaining values $\hat{u}(\xi(\lambda,\varphi))$
on the polar grid;

\item Use the bi-linear interpolation to obtain values $\hat{u}(\xi)$ on a
Cartesian grid from values $\hat{u}(\xi(\lambda,\varphi))$:

\item Using the 2D FFT reconstruct $u(x)$ from $\hat{u}(\xi)$;

\item Compute $[\mathcal{A}^{\ast}g](x)$ as $\left[  \Pi_{\Omega_{0}}u\right]
(x).$
\end{algorithmic}
\end{algorithm}

Similarly to our Algorithm \ref{alg:forward}, 
the most time consuming  steps of the present technique are the computation of 2D FFT which requires
$\mathcal{O}(n^{2}\log n)$ flops, and $\mathcal{O}(n)$ evaluations of 1D
FFT's, which results in the $\mathcal{O}(n^{2}\log n)$ flops count for the
whole algorithm. Additional time saving is achieved by pre-computing and
storing the values of the Bessel functions $J_{k}(\lambda)$ for the values of
$k$ from $0$ to half the number of discretization points in $\theta$, and
for all values of $\lambda=|\xi|$ in the selected polar grid in $\xi$.

\subsection{Inverse operator}

Although the main goal of this paper is the development of fast algorithms for
computing $\mathcal{A}$ and $\mathcal{A}^{\ast}$, the approach used for fast
evaluation of $\mathcal{A}^{\ast}$ can be easily modified to obtain a fast
Hankel-function-free algorithm for approximating the inverse $\mathcal{A}
^{-1}$ defined by the universal backprojection formula in 2D, equation
(\ref{E:inverse}). Indeed, while formula (\ref{E:adjoint}) can be interpreted
as a time derivative of the solution of the free space wave equation with a
single layer potential supported on $S,$ the backprojection formula equation
(\ref{E:inverse}) can be understood as a computation of a double layer
potential also supported on $S$. Formula (\ref{E:inverse}) yields a
theoretically exact reconstruction of $f$ from $g=\mathcal{A}f$ under the
condition that the data $g$ are given on the time interval $(0,\infty)$. If
the data are given of on the time interval $(0,T)$ with $T>2$, the
reconstruction will contain an infinitely smooth error. In most practical
cases such an error is dominated by other imperfections of measurements, so
that an acceptable approximation to $f$ can be computed as
\begin{align}
f(x)  &  =[\mathcal{A}^{-1}g](x)\approx\left[  \Pi_{\Omega_{0}}v\right]
(x),\nonumber\\
v(x)  &  \equiv2\int\limits_{0}^{T}\int\limits_{\Omega}g(t,z)\frac{\partial
}{\partial n(z)}G(t,x-z)\,dz\,dt,\qquad x\in R^{2}. \label{E:approx_UBP}
\end{align}
The restriction operator $\Pi_{\Omega_{0}}$ appears in the above formula since
the function $v(x)$ we compute is supported in the square $[-L,L]\times
\lbrack-L,L]$ with $L=1+T$

A computation similar to the one done in the previous section allows one to
obtain for the 2D Fourier transform $\hat{v}(\xi)$ of the function $v(x)$ the
following expression:
\begin{align}
\hat{v}(\xi)  &  =\sum_{k=-\infty}^{\infty}e^{ik\varphi}\hat{v}_{k}
(\lambda),\label{E:Four_ser_v}\\
\hat{v}_{k}(\lambda)  &  \equiv-4\pi(-i)^{|k|}J_{|k|}^{\prime}(\lambda
)\int\limits_{0}^{T}g_{k}(t)\sin(\lambda t)\,dt, \label{E:Four_coef_v}
\end{align}
where the Fourier coefficients $g_{k}(t)$ are still given by
(\ref{E:FourSerForg}) (we relegate the details of this computation into the Appendix).

For the algorithm approximating $\mathcal{A}^{-1}$, we use the same sizes of
grids and computational domains as for $\mathcal{A}^{\ast}$, so that the
resulting technique is a small modification of the latter algorithm. An
additional step resulting in a significant reduction of the error due to
finite-time data we use is an addition of a constant $C$ chosen so that
\begin{equation}
\int\limits_{\Omega\backslash\Omega_{0}}[v(x)+C]dx=0. \label{E:silly_integral}
\end{equation}
Here we are just using the $\emph{apriori}$ information that $f(x)$ is
supported in $\Omega_{0}$.
This results in Algorithm \ref{alg:inverse}.

\begin{algorithm}
\caption{Computing $\mathcal{A}^{-1}g$}
\label{alg:inverse}
\begin{algorithmic}[1]
\item Extend $g(t,x)$ by zero from $(0,T)\times\Gamma$ to
$(0,T_{\mathrm{large}})\times S$;

\item Using the 1D FFTs, compute coefficients $g_{k}(t)$ (equation
(\ref{E:FourSerForg})) for each grid value of $t$;

\item Using the 1D sine FFTs, compute coefficients $\hat{v}_{k}(\lambda)$ for
each grid value of $\lambda$, equation (\ref{E:Four_coef_v});

\item Using the 1D FFTs, sum the Fourier series (\ref{E:Four_ser_v}) for each
grid value of $\lambda$, thus obtaining values $\hat{v}(\xi(\lambda,\varphi))$
on the polar grid;

\item Use the bi-linear interpolation to obtain values $\hat{v}(\xi)$ on a
Cartesian grid from values $\hat{v}(\xi(\lambda,\varphi))$:

\item Using the 2D FFT reconstruct $v(x)$ from $\hat{v}(\xi)$;

\item Find constant $C$ to satisfy (\ref{E:silly_integral});

\item Compute an approximation to $[\mathcal{A}^{-1}g](x)$ as $\left[
\Pi_{\Omega_{0}}(v(x)+C)\right]  (x).$
\end{algorithmic}
\end{algorithm}
Similarly to our Algorithm \ref{alg:adjoint} for $\mathcal{A}^{\ast}$, the present technique
requires $\mathcal{O}(n^{2}\log n)$ flops.
\begin{figure}[t]
\begin{center}
\subfigure[Computed $g=\mathcal{A}f$]
{\includegraphics[scale=0.45]{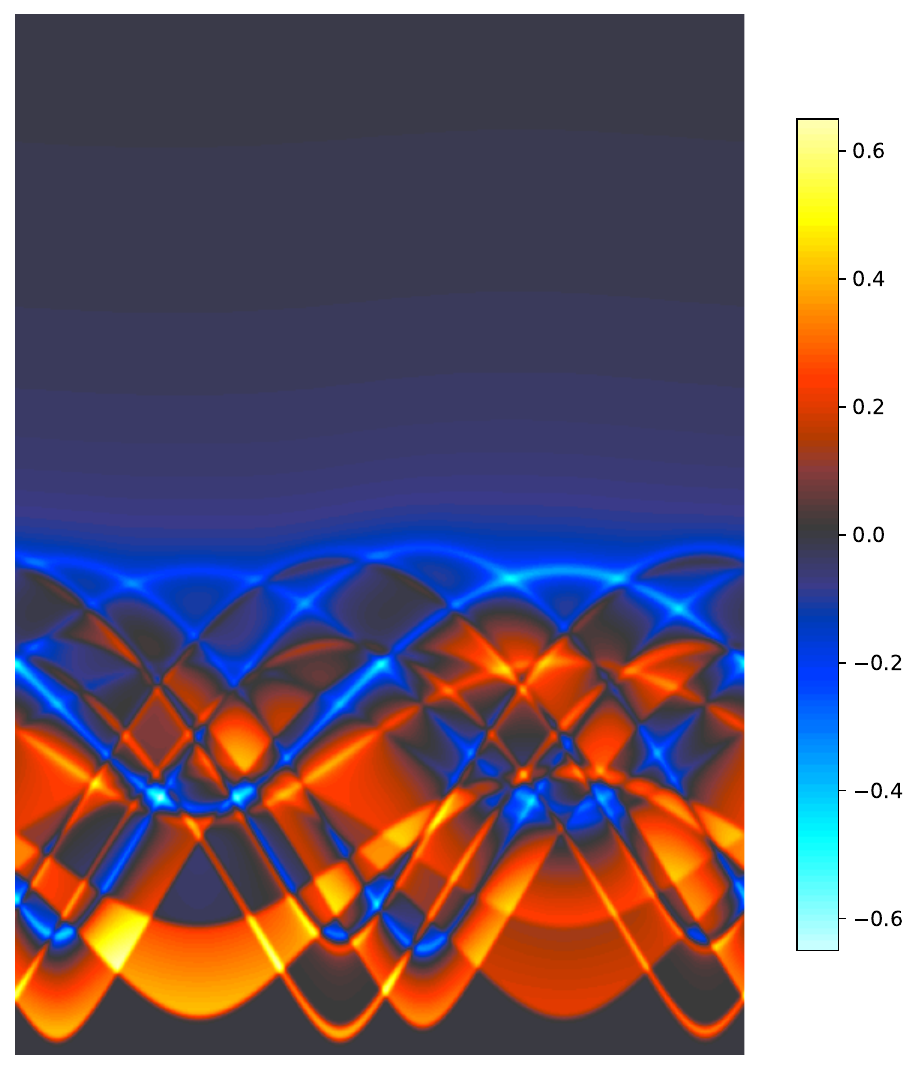}}
\subfigure[Error in the computed $g$]
{\includegraphics[scale=0.45]{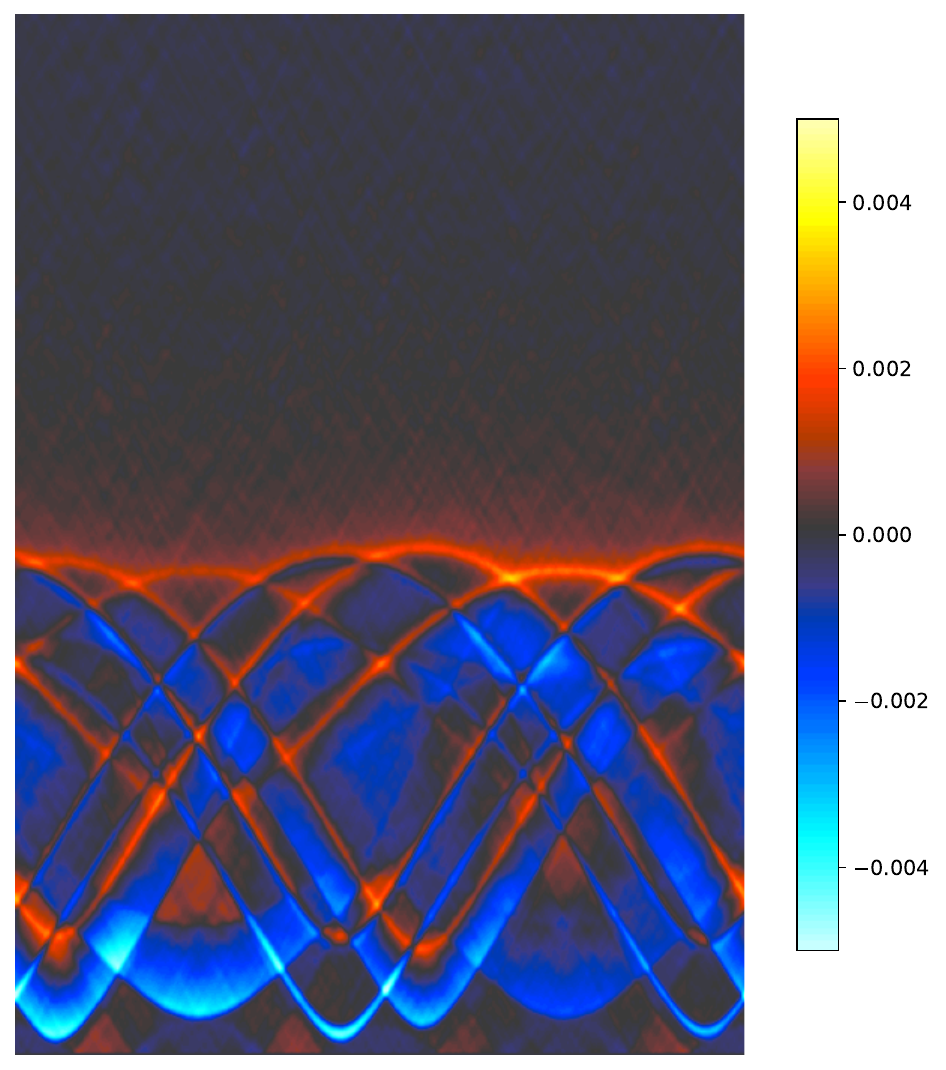}}
\end{center}
\par
\vspace{-0.5cm}\caption{Reconstruction error in the fast forward computation}
\label{F:forward}
\end{figure}

\section{Numerical simulations}\label{sec:numericalEval}

In this section, we test our algorithms in a set of numerical simulations,
covering different subsets of circular acquisition geometry. In particular,
we investigate: (a) the complete data geometry, where surface $\Gamma$
coincides with the circle $S$; (b) the 180 degree acquisition with the
support of $f$ lying inside the upper half of the circle $S$ in such a way
that the whole object is "visible" (see~\cite{KuKuEncycl} for the discussion of visible
singularities); and (c) the 180 and 120 degrees acquisition with the support
of $f$ occupying most of the interior of $S,$ leading to a wide set of
"invisible" singularities.

\subsection{Forward problem}

First, we tested our algorithm for computing the result of application of the
forward operator $\mathcal{A}$ to a given function $f.$ Here and further in
the test (except Section~\ref{S:half_image}) we assume that support $\Omega_{0}$ of $f$ is a
disk of radius 0.98 centered at the origin. Our phantoms $f$ are modeled by a
rather arbitrary set of smoothed characteristic functions of circles. For the
example presented below, function $f$ was discretized on a $257\times257$
Cartesian grid covering square $[-1,1]\times\lbrack-1,1]$; the color map
representation of this function is shown in figure \ref{F:360clean}(a). The
solution to the forward problem $g=\mathcal{A}f$ was computed using the algorithm
described in Section \ref{S:forward}, on a computational grid with $360$
angular positions and $513$ uniformly spaced points in time covering interval
$[0,4].$ All of our simulations were run on an Nvidia A2000 GPU. The
computation time for one application of the forward operator $\mathcal{A}$
took $0.0055$ sec (averaged over 100 runs), not counting the time spent
precomputing the Bessel functions. The computed $g$ is shown in Figure
\ref{F:forward}(a). We compared our result to $\mathcal{A}f$ calculated using
a simplified version of the \emph{k-Wave} algorithm \cite{kwave} and a
finer sampled version of $f$ (on $513\times513$ grid). The difference (which
can be considered the error in our method) is plotted in Figure
\ref{F:forward}(b), using a much finer color scale. The relative $L_{2}$ error
was 0.58\%, and the relative $L^{\infty}$ error was 0.8\%.

\subsection{Reconstruction methods}\label{sec:ReconMethods}

In the following we will compare different reconstruction methods using the introduced fast FFT models to illustrate their usefulness for solving common reconstruction approaches. The first method will be simply the application of the FFT inverse outlined in Alg. \ref{alg:inverse}. This is expected to provide good reconstructions for full-view data and where the visibility condition is satisfied. However, for limited-view and noisy data the inverse $\mathcal{A}^{-1}$ is not expected to provide satisfactory results and hence we will additionally present results for two iterative methods, non-negative least squares (NNLS) as well as total variation (TV) regularization. Both approaches compute a minimizer by solving the variational problem in \eqref{E:minim}. Both methods need to be solved iteratively and benefit strongly from the fast FFT forward in Alg. \ref{alg:forward} and adjoint in Alg. \ref{alg:adjoint}. We will provide a short summary below for each method.

The first approach, NNLS, solves only the least squares problem with non-negativity, i.e.,
\begin{equation}\label{eqn:NNLS}
\argmin_{f\geq 0} \|\mathcal{A}f-g\|_{2}^{2}.
\end{equation}
This can be efficiently implemented as a projected gradient descent, or proximal gradient descent for the non-negativity constraint, as follows
\begin{align}
{f}^{(k+\frac{1}{2})}  &  =f^{(k)} -  \lambda \mathcal{A}^{\ast}\left(\mathcal{A}f^{(k)} - g \right),\\
f^{(k+1)}  &  =\max(0,{f}^{(k+\frac{1}{2})}), \qquad k=0,1,2, \dots, \label{eq:NNLS-projstep}
\end{align}
with initialization $f^{(0)}=0$. The projection step \eqref{eq:NNLS-projstep} can additionally contain a projection to a region of interest $\Pi_{\Omega_{0}}$, which will be utilized in the case where we only reconstruct in the upper half of the unit disk $D$ where the visibility condition is satisfied. In this case the full projection step is given by
\begin{equation}\label{eqn:projectionHalfCircle}
f^{(k+1)}  =\Pi_{\Omega_{0}}\max(0,{f}^{(k+\frac{1}{2})}).
\end{equation}

While NNLS can mitigate limited-view artifacts, it is not able to remove noise. Thus, we will additionally employ the well-established TV regularization, which adds the regularizer in \eqref{E:minim} penalizing the gradient of $f$, that is
\[
\mathcal{R}(f) = \| |\nabla f| \|_1,
\]
here the 1-norm enforces sparsity in the gradient and hence promotes piecewise constant reconstructions. Since the regularizer is non-differentiable, we will solve the TV problem with the the primal-dual hybrid gradient (PDHG) method \cite{chambolle2011first}, which allows for convex non-differentiable $\mathcal{R}$. The algorithm can be stated for the problem \eqref{E:minim} with TV regularization by
\begin{align}
    q^{(k+1)} &=\frac{q_k + \sigma (\mathcal{A} \widetilde{f}^{(k)} - g)}{1+\sigma} \label{eqn:PDHG-dual}, \\
    f^{(k+1)} &=\text{prox}_{\mathcal{R},\alpha\lambda}\left(f^{(k)} - \lambda \mathcal{A}^* q^{(k+1)}\right), \label{eqn:PDHG-primal} \\
    \widetilde{f}^{(k+1)}&= f^{(k+1)} + \mu(f^{(k)} - f^{(k+1)}), \quad k=0,1,2,\dots . \label{eqn:PDHG-relaxation}
\end{align}
where $\sigma,\lambda>0$, $\mu\in[0,1]$, regularization parameter $\alpha>0$, and initializations $f^{(0)}=0$ and $q^{(0)}=0$.
The proximal operator in \eqref{eqn:PDHG-primal} enforces the regularizer by solving the corresponding denoising problem
\[
\text{prox}_{\mathcal{R},\alpha\lambda}(f) = \argmin_{h} \left\lbrace \mathcal{R}(h) + \frac{1}{2 \alpha\lambda} \|h - f\|_2^2 \right\rbrace.
\]
Note, that this can also be understood as a projection step to the space of permissible solutions with respect to the regularizer $\mathcal{R}$.
Let us point out, that \eqref{eqn:PDHG-dual} corresponds to the proximal operator in the dual space, i.e., enforcing the least-squares data fidelity term.

\subsubsection{Learned primal dual reconstructions}

Additionally, we will test the capabilities of the forward/adjoint models for training a deep learning model, and in particular an iterative learned image reconstruction method, which makes use of the forward and adjoint operator in the network architecture. Specifically, we will test the learned primal dual (LPD) method \cite{Adler2018} which reformulates the PDHG above using neural networks instead of the proximal operators \eqref{eqn:PDHG-dual} and \eqref{eqn:PDHG-primal}. That is, we introduce two convolutional neural networks $\Gamma_{\phi}$ and $\Lambda_{\psi}$ which operate on the dual (measurement space) and primal (image space), respectively. The LPD algorithm then reformulates the updates in \eqref{eqn:PDHG-dual} and \eqref{eqn:PDHG-primal} to a learned version as follows
\begin{align}
    q^{(k+1)} &=\Gamma_{\phi}\left(q^{(k)}, \mathcal{A} f^{(k)}, y\right) \label{eq:LPD_dual} \\
    f^{(k+1)} &=\Lambda_{\psi}\left(f^{(k)}, \mathcal{A}^* q^{(k+1)}\right), \quad k=0,1,2,\dots . \label{eq:LPD_primal}
\end{align}
Given a finite number of iterations $K>0$, the above update rules define a learned reconstruction operator $\mathcal{A}^ \dagger_{\phi,\psi}\colon y \mapsto f^K$ with learnable parameters $\psi$ and $\phi$. Note, that this uses weight-sharing, i.e., each iterate has the same parameters. Despite the weight-sharing these can be high dimensional, in our case the full reconstruction operator $\mathcal{A}^\dagger_{\phi,\psi}$ has around $3.9\cdot 10^6$ learnable parameters.

This reconstruction operator $\mathcal{A}^ \dagger_{\phi,\psi}$ can now be trained given appropriate training data. Here, we consider paired supervised training data $\{f_i,g_i\}_{i=1}^N$, which satisfy the operator equation
\begin{equation}\label{eq:Oper_equation}
    \mathcal{A}f + \delta = g,
\end{equation}
with noise $\delta$. The reconstruction operator is then trained by minimizing the empirical loss to find optimal parameters $(\psi^*,\phi^*)$ by
\begin{equation}
(\psi^*,\phi^*) = \argmin_{(\psi,\phi)} \sum_{i=1}^N \|\mathcal{A}^ \dagger_{\phi,\psi}(g_i) - f_i\|_2^2. \label{eq:emp_loss}
\end{equation}
Note, that the minimization of \eqref{eq:emp_loss} requires to first evaluate $\mathcal{A}^ \dagger_{\phi,\psi}$ (the forward pass) which involves evaluating $\mathcal{A}$ and $\mathcal{A}^*$ each $K$-times. Additionally, computing the gradients with respect to the parameters $(\phi,\psi)$ requires to differentiate through \eqref{eq:LPD_dual} and \eqref{eq:LPD_primal}, leading two $K$-times additional evaluations of forward and adjoint. That is, in total $4K$ operator evaluations per training iteration. This clearly necessitates efficient implementations.

\begin{figure}[t]
\begin{center}
\subfigure[Measurement geometry with 180 deg.]
{\includegraphics[scale=0.50]{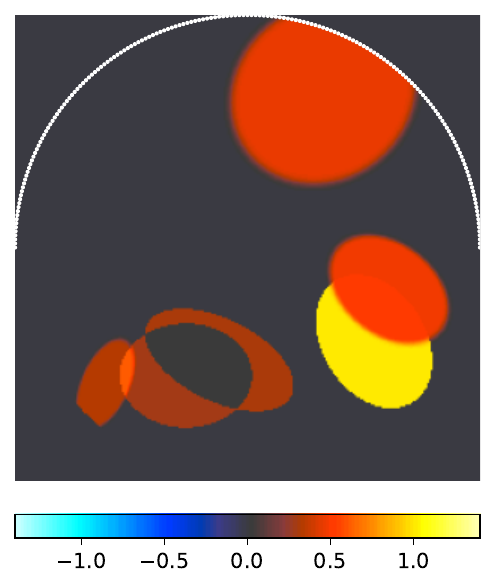}
\includegraphics[scale=0.50]{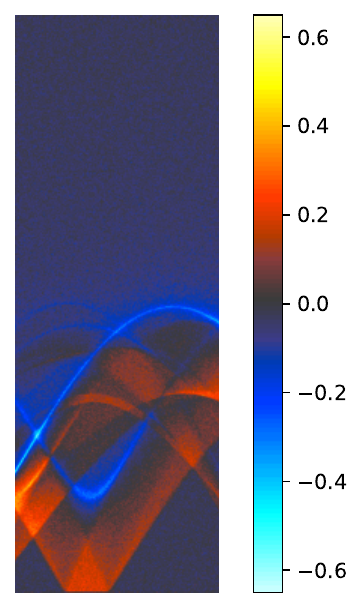}}
\subfigure[Measurement geometry with 120 deg.]
{\includegraphics[scale=0.50]{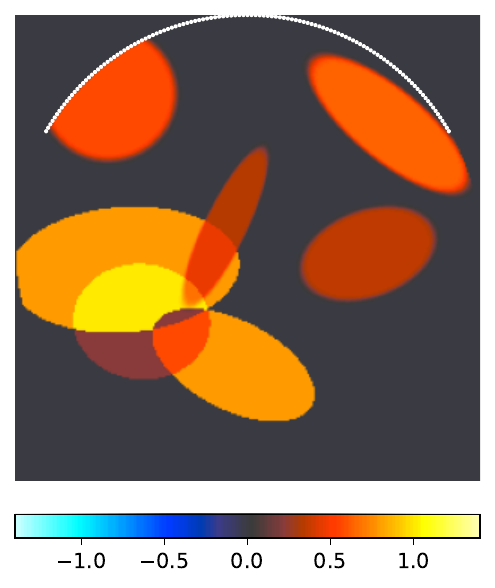}
\phantom{a}\includegraphics[scale=0.50]{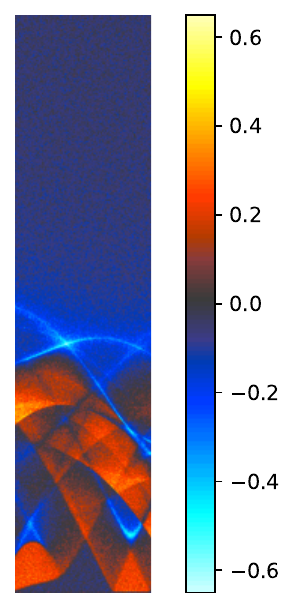}}
\end{center}
\par
\vspace{-0.5cm}\caption{Two samples (a) and (b) from the training data with each ground truth $f$ (left) and computed $g=\mathcal{A}f + \delta$ (right). The dotted white lines indicate the location of the transducers. }
\label{F:trainingdata}
\end{figure}

For this study, we have chosen the number of LPD iterations $K = 10$ and a U-Net type architecture for each of the networks $\Gamma_\theta$ and $\Lambda_\psi$. We used the same architecture for both primal and dual network, apart from different input dimensions as outlined in \eqref{eq:LPD_primal} and \eqref{eq:LPD_dual}. The used U-Net consisted of four scales, i.e. three down and up-sampling layers, with a window size of 2. In each scale, we applied two convolutional layers followed by batch normalization and ReLU. We have chosen the same size of all convolutional kernels as 3 × 3 and the width in the first scale 32, which in each downsampling was doubled, that is in the finest scale we used 256 filters.  We trained the LPD networks by using the ADAM optimizer with a cosine decay on the learning rate initialized as $10^{-4}$ for a total of  100 000 training iterations.  The training data consisted of \mbox{10 000} phantoms of randomly generated ellipses with smoothed boundaries of varying degree and we used 30\% Gaussian noise in the operator equation \eqref{eq:Oper_equation}. Examples from the training data are shown in figure \ref{F:trainingdata}. Testing was performed on the same phantom as for all other methods.

\subsection{Inverse problem with complete data}\label{S:complete}

\begin{figure}[t]
\begin{center}
\subfigure[Both ground truth $f$ and $\mathcal{A}^{-1} g$ ]
{\includegraphics[scale=0.4]{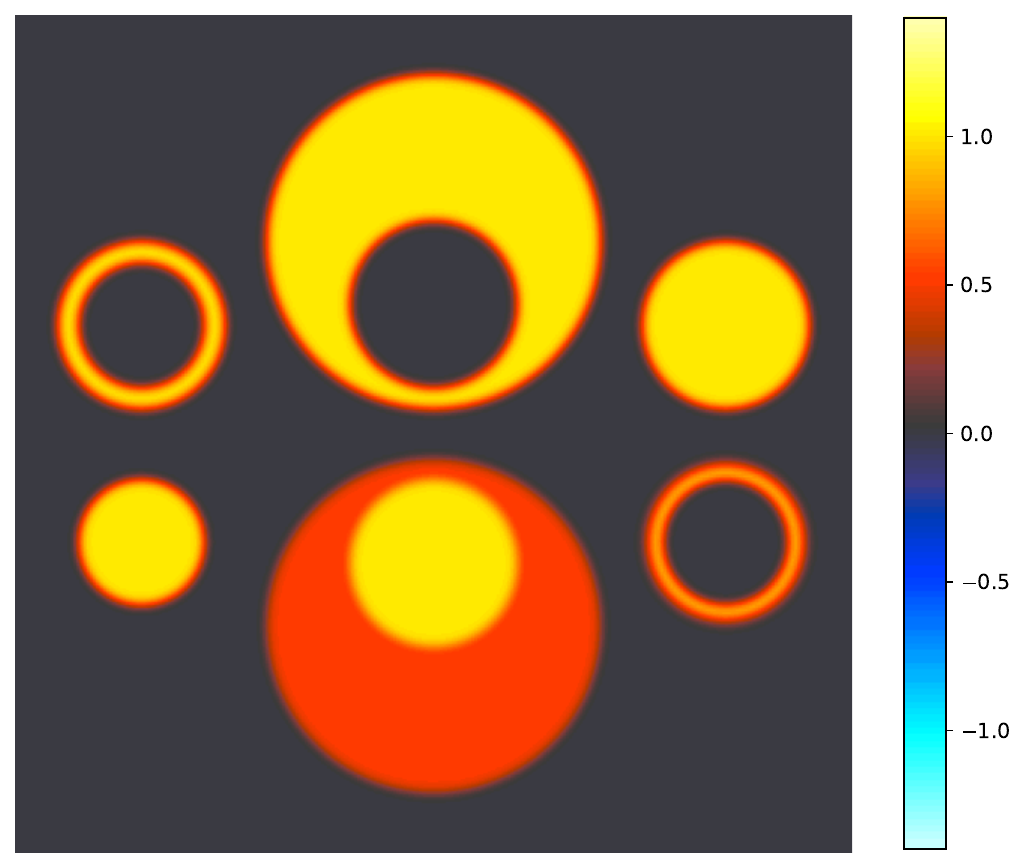}}
\subfigure[Error in $\mathcal{A}^{-1} g$]
{\includegraphics[scale=0.4]{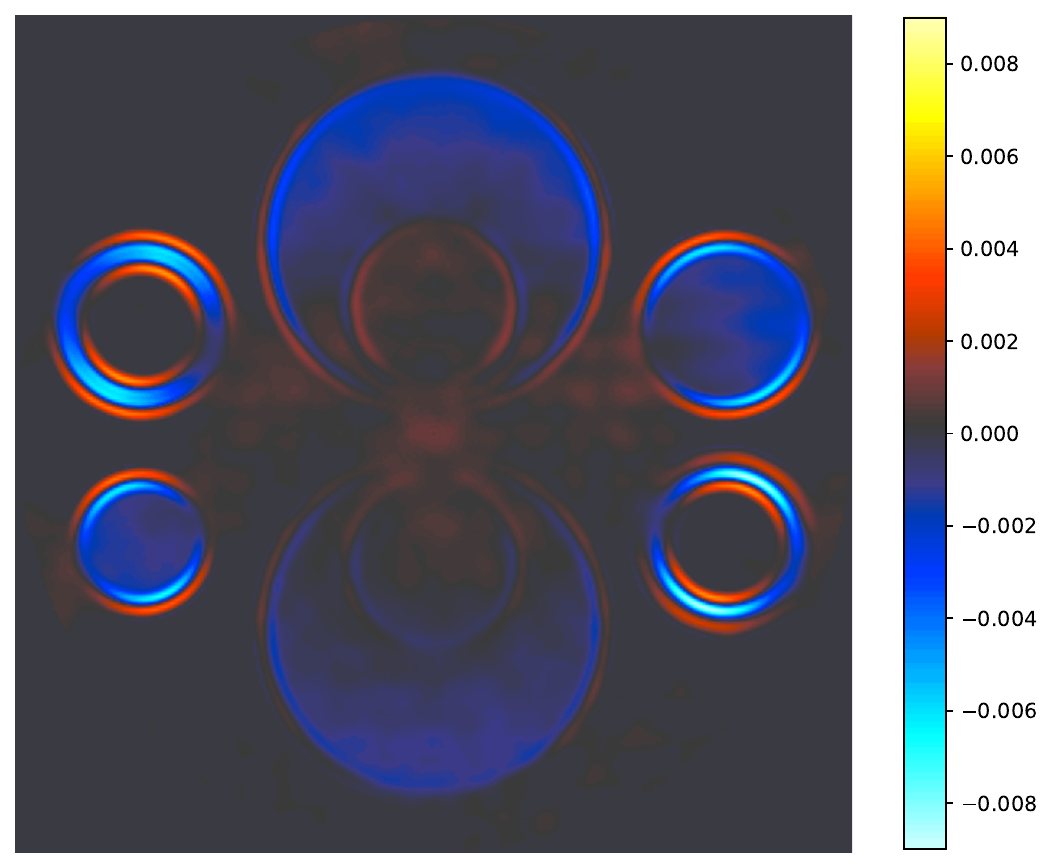}}
\subfigure[Finite difference reconstruction ]
{\includegraphics[scale=0.4]{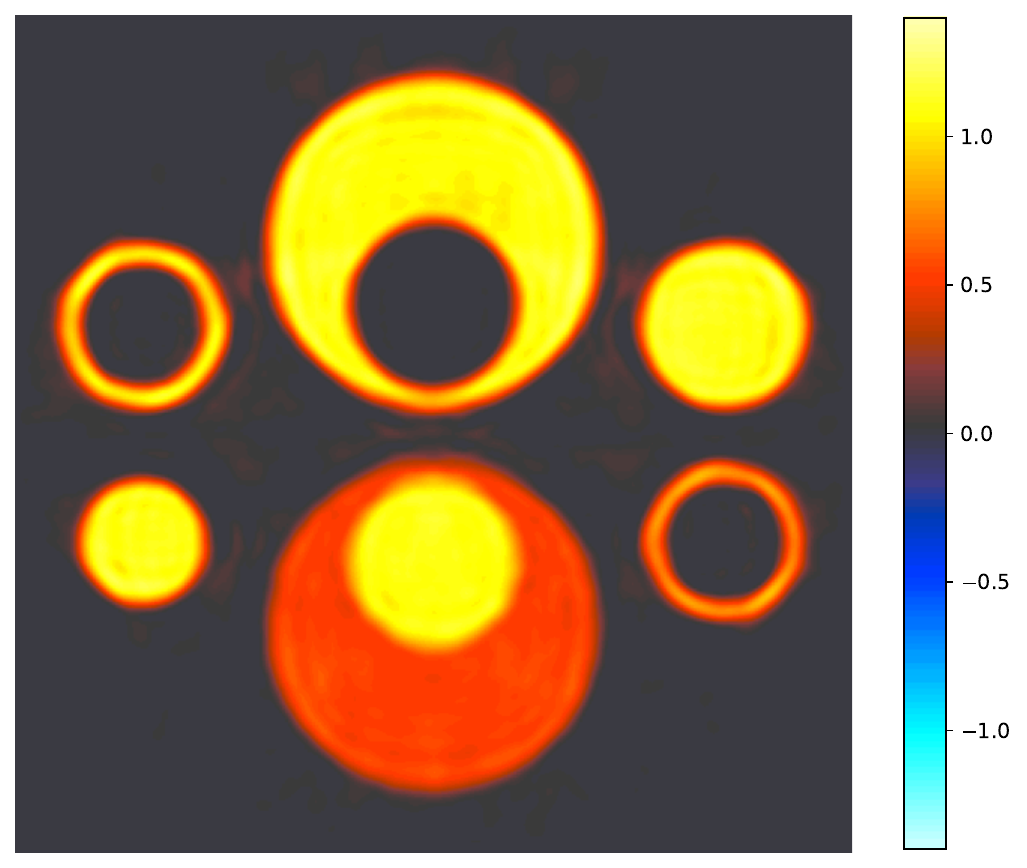}}
\subfigure[Error in fin diff image]
{\includegraphics[scale=0.4]{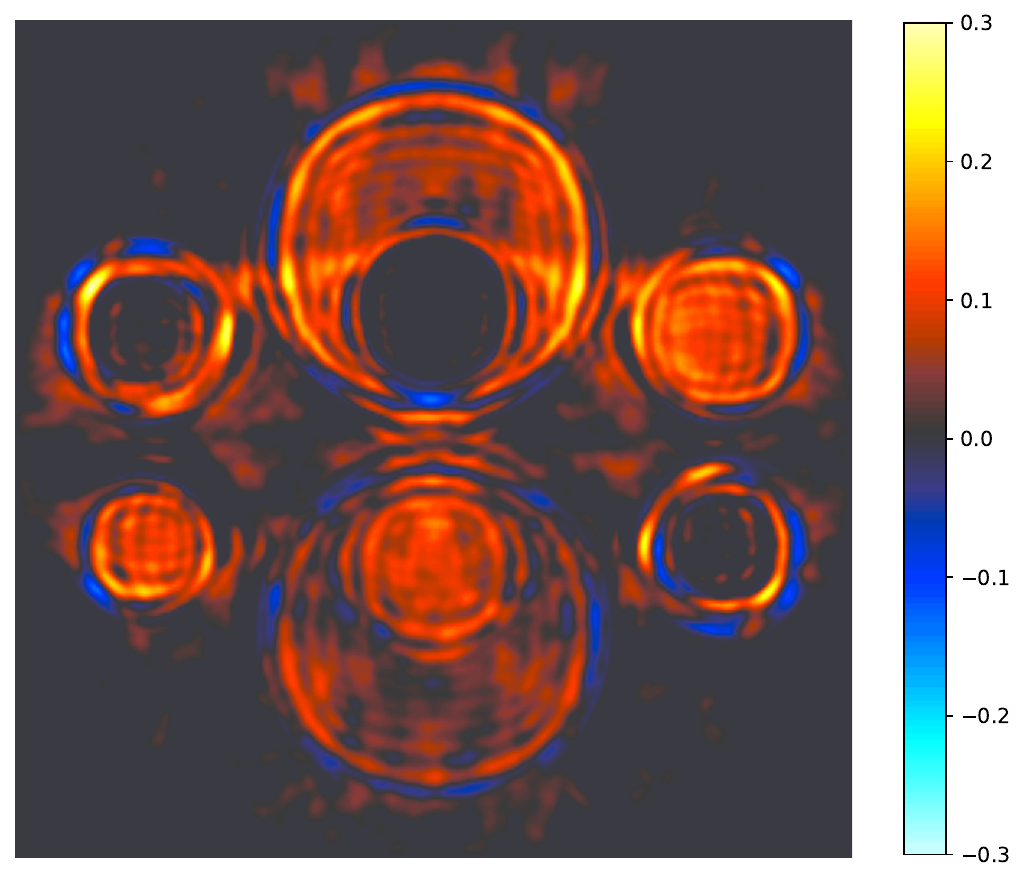}}
\end{center}
\par
\vspace{-0.5cm}\caption{360 deg acquisition, accurate data}
\label{F:360clean}
\end{figure}

Here we assume that support $\Omega_{0}$ of $f$ is a disk of radius 0.98
centered at the origin, and that the data $g=\mathcal{A}f$ are collected on a
time-space cylinder $[0,4]\times S.$ As the data $g$ we used the accurate
approximation computed in the previous section using the slow \emph{k-Wave}
algorithm on the grid $513\times360$, using finely discretized version of the
phantom $f$ shown in figure \ref{F:360clean}(a). The reconstructed image
$\mathcal{A}^{-1}g$ obtained by our inverse algorithm is indistinguishable from
the ground truth $f$ when plotted as a color map. The error of the
reconstruction is shown in figure \ref{F:360clean}(b) on a much finer color
scale. The relative error in $L_{2}$ norm is 0.22\%, and the relative
$L^{\infty}$ error is 0.9\%. The reconstruction time (not counting
the precomputation of the Bessels functions) was 0.011 sec; it was obtained by
averaging one hundred computations of $\mathcal{A}^{-1}$. For comparison, we
also computed an approximation to $f$ using the finite difference time reversal method
\cite{hkn,BurgPhysRev} mentioned in Section \ref{S:known_forward}.
This
technique has the complexity of $\mathcal{O}(n^{3})$ flops, with a small
hidden constant due to its simplicity. We implemented this algorithm using the
Torch package and ran it on an Nvidia A2000 GPU. The number of time steps had
to be doubled for this technique, due to the stability requirement for the
explicit finite difference time stepping of this method. The computation time
was $0.19$ seconds. This is at least an order of magnitude slower than our
fast algorithm. More importantly, this method produced reconstruction of a
much inferior quality. The reconstructed image is shown in figure
\ref{F:360clean}(c) and the corresponding error is plotted in \ref{F:360clean}
(d), on a different color scale. The relative reconstruction error was is 11\%
in $L_{2}$ norm 30\% in $L^{\infty}$ norm.

\begin{figure}[t]
\begin{center}
\subfigure[Reconstruction via $\mathcal{A}^{-1} g$, 30\% noise in $g$]
{\includegraphics[scale=0.4]{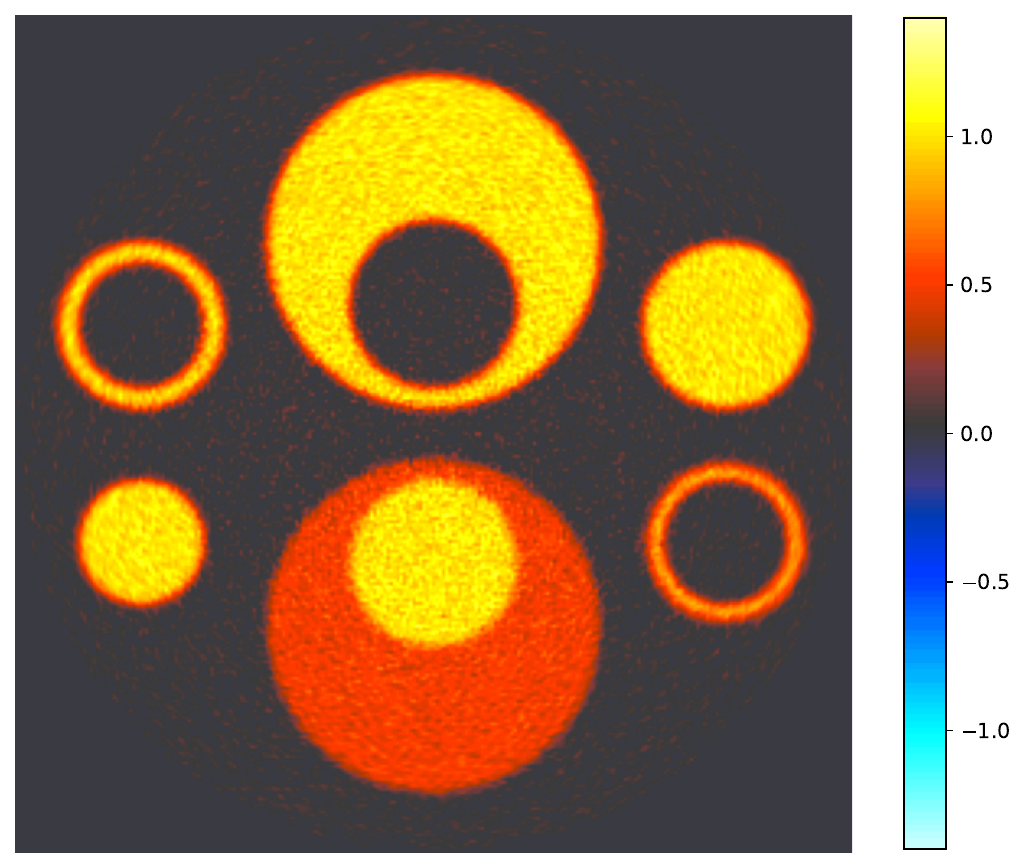}}
\subfigure[Iterative TV reconstruction, 30\% noise in $g$]
{\includegraphics[scale=0.4]{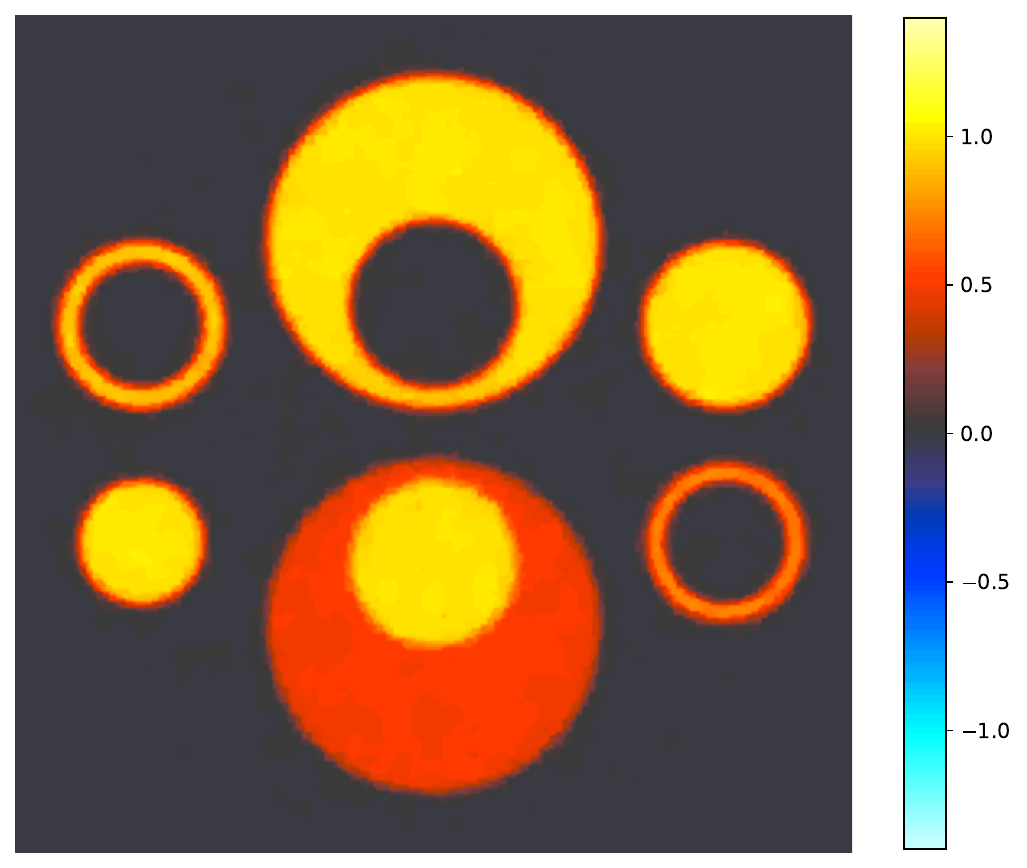}}
\end{center}
\par
\vspace{-0.5cm}\caption{360 deg acquisition, 30\% noise in the data}
\label{F:360noise}
\end{figure}

In order to simulate significant measurement noise present in real application
of PAT, TAT\ and other hybrid imaging modalities, we added to the data white
noise, computed as a sequence of realizations of a normally distributed random
variable. The relative intensity of the noise measured in $L_{2}$ norm was 30\%.
The image reconstructed by application of our fast inverse algorithm without
any additional regularization is shown in figure \ref{F:360noise}(a). As one
would expect, the reconstruction is quite noisy. The relative error measured
in $L_{2}$ norm is 9.9\%, while in $L^{\infty}$ norm it is 30\%. An improved
reconstruction was calculated using the iterative algorithm with TV
regularization as described in Section \ref{sec:ReconMethods}.
For this algorithm, and for all other iterative algorithms described below we used the
following stopping criterion: the $L_{2}$ norm of an update to the current
approximation is smaller than 0.3\% of the $L_{2}$ norm of the very first
(non-zero) approximation. The method took 53 iterations to converge, with a total computation time of 2.8 seconds (including precomputation of Bessel
functions). The resulting image is shown in Figure \ref{F:360noise}(b). The
relative $L_{2}$ norm of the error was reduced to 5.5\% and the relative
$L^{\infty}$ norm was found to be 22\%.

\subsection{Inverse problem with partial data}

\begin{figure}[t]
\begin{center}
\subfigure[Ground truth $f$ ]
{\includegraphics[scale=0.362]{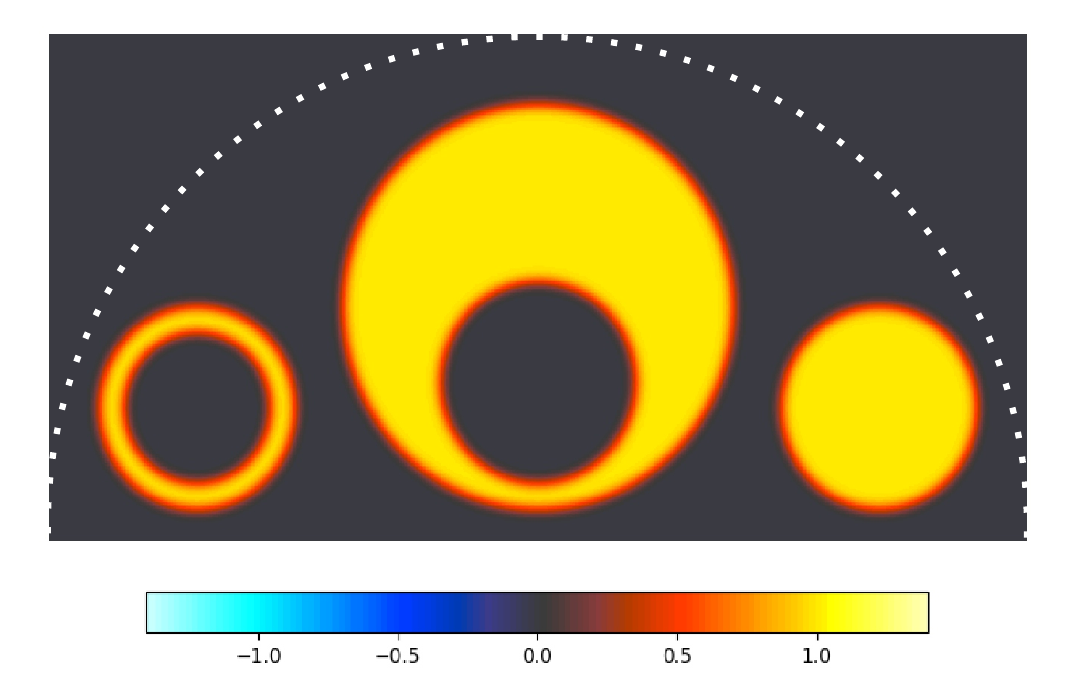}}
\subfigure[The inverse $\mathcal{A}^{-1} g$, no noise in $g$]
{\includegraphics[scale=0.35]{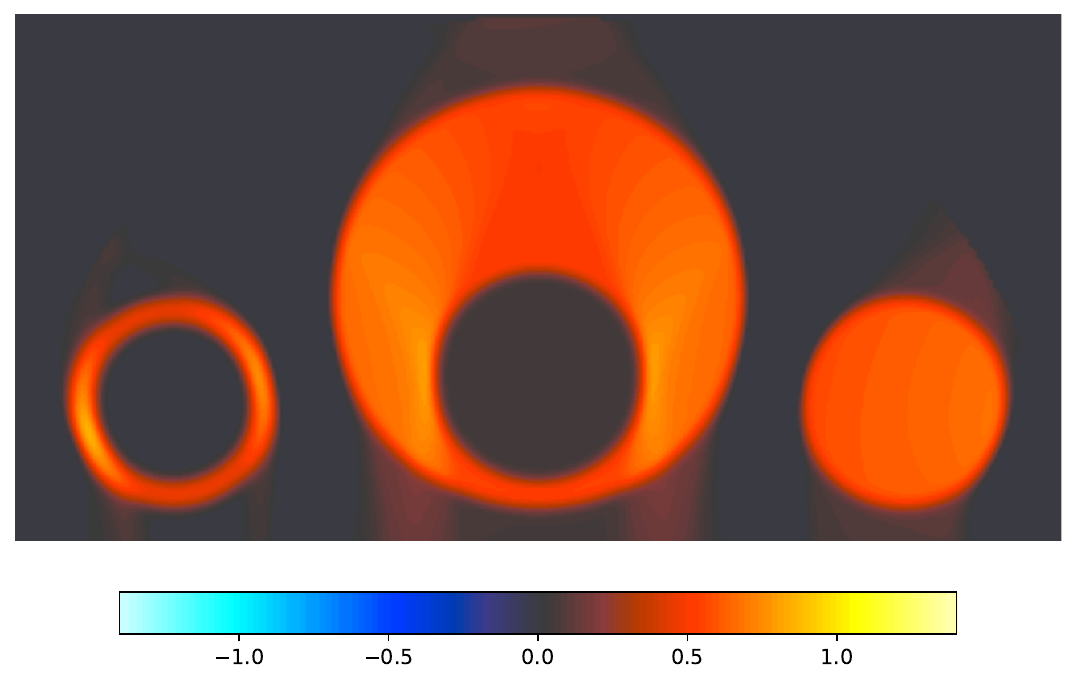}}
\subfigure[NNLS iterations, 30\% noise]
{\includegraphics[scale=0.35]{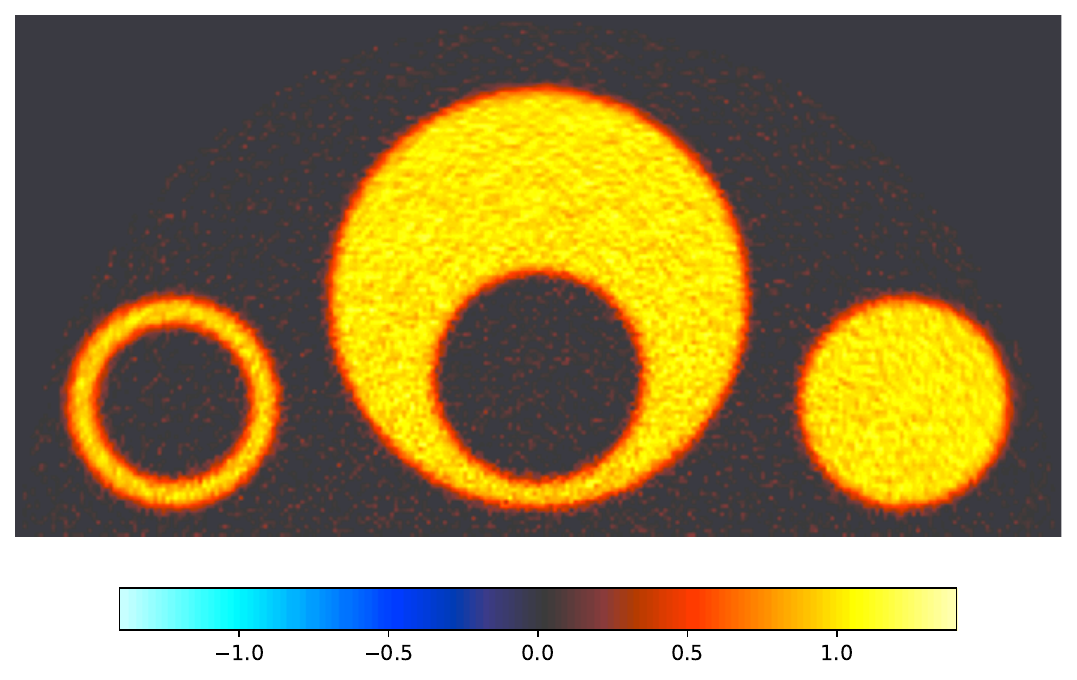}}
\subfigure[Iterative TV, 30\% noise in $g$]
{\phantom{a}\includegraphics[scale=0.35]{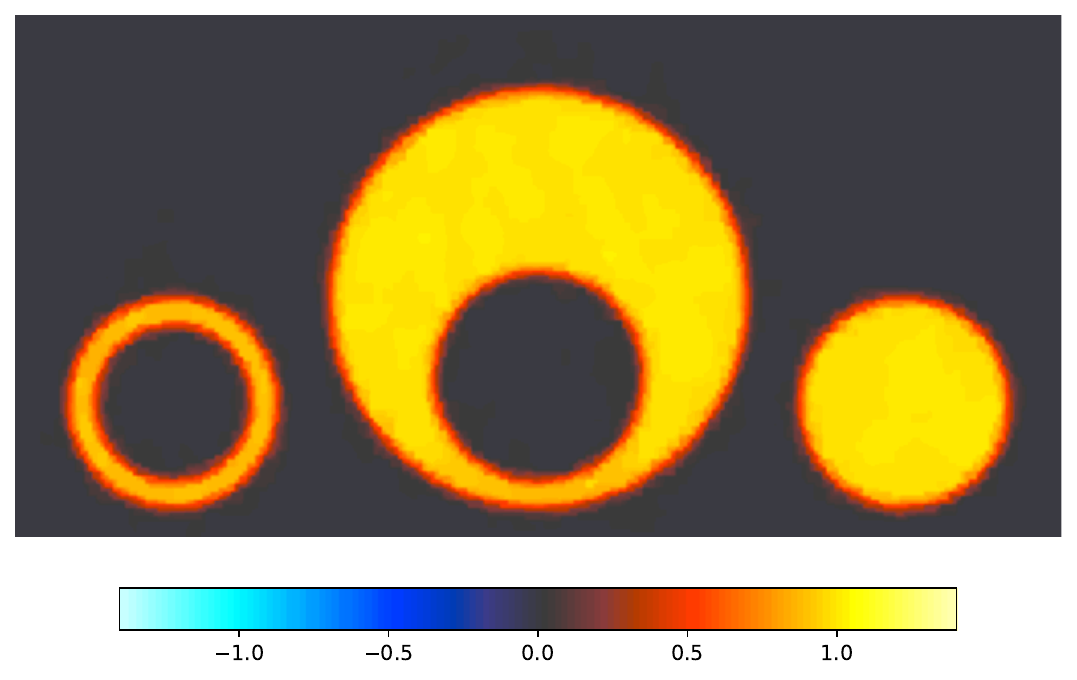}}
\end{center}
\par
\vspace{-0.5cm}\caption{180 deg acquisition, with a support satisfying
visibility condition, 30\% noise. Dotted white line in the part (a) shows the
location of the transducers.}
\label{F:180half}
\end{figure}

In this Section, we consider an inverse problem in which data are measured on
the upper half of the circle $S$, and the support of $f$ is the upper-half of
the disk $D\equiv\{x\in\mathbb{R}^{2}:|x|<0.98\}$. This measuring scheme
satisfies the "visibility condition" (see, e.g. \cite{KuKuEncycl}), and the
problem of reconstructing $f$ in such geometry is well-posed. As a phantom we
used the upper half of the phantom used in the previous section, as shown in
Figure \ref{F:180half}(a). The white dotted line in this figure indicates the
location of the transducers. The data $g$ was computed as in section
\ref{S:forward}, with values corresponding to the lower half of the circle
reset to $0.$ The application of our fast inverse operator yields an image
with significant artifacts, see Figure \ref{F:180half}(b). The relative
$L_{2}$ error in this reconstruction is 40\%, with $L^{\infty}$ norm equal
to 51\%.

However, due to the ill-posedness of this problem, a much better
image can be obtained by variational techniques as previously discussed in Section \ref{sec:ReconMethods}. For solving the NNLS problem \eqref{eqn:NNLS}, we will use the additional projection step in \eqref{eqn:projectionHalfCircle},  where $\Pi_{\Omega_{0}}$ is the restriction of a function to the upper half of the disk $D$.
This computation converged in 17 iterations that took 2
seconds (with the same stopping criterion as before). The color map image of
the reconstructed approximation to $f$ is indistinguishable from the ground
truth shown in Figure \ref{F:180half}(a). The relative $L_{2}$ error in this
reconstruction is 0.5\%, while the $L^{\infty}$ norm of the error is 2.8\%.

The images in figures \ref{F:180half}(c) and \ref{F:180half}(d) present the
reconstructions from the same data but with added 30\% white noise (as
measured by the relative $L_{2}$ norm). The image in \ref{F:180half}(c) was
obtained by NNLS. The algorithm converged in 26 iterations that
took 2.2 sec. The relative $L^{2}$ error in this image is 11\%, the
$L^{\infty}$ norm of the error is 37\%. The reconstruction depicted in
\ref{F:180half}(d) was obtained by the iterative algorithm with TV
regularization. It took 74 iterations (or 3 seconds) to converge. The relative
$L_{2}$ error got reduced to 5.2\% while the relative $L^{\infty}$ norm of the
error was 26\%. \begin{figure}[t]
\begin{center}
\subfigure[The inverse $\mathcal{A}^{-1} g$]
{\includegraphics[scale=0.4]{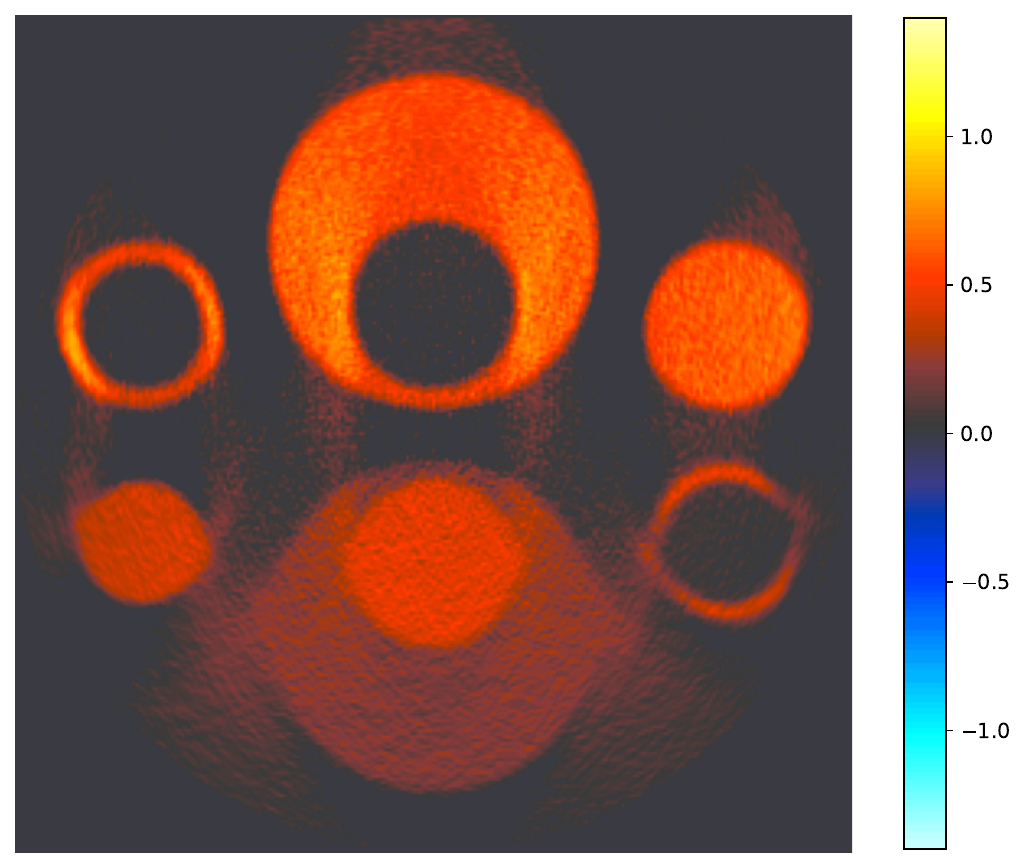}}
\subfigure[NNLS iterations]
{\includegraphics[scale=0.4]{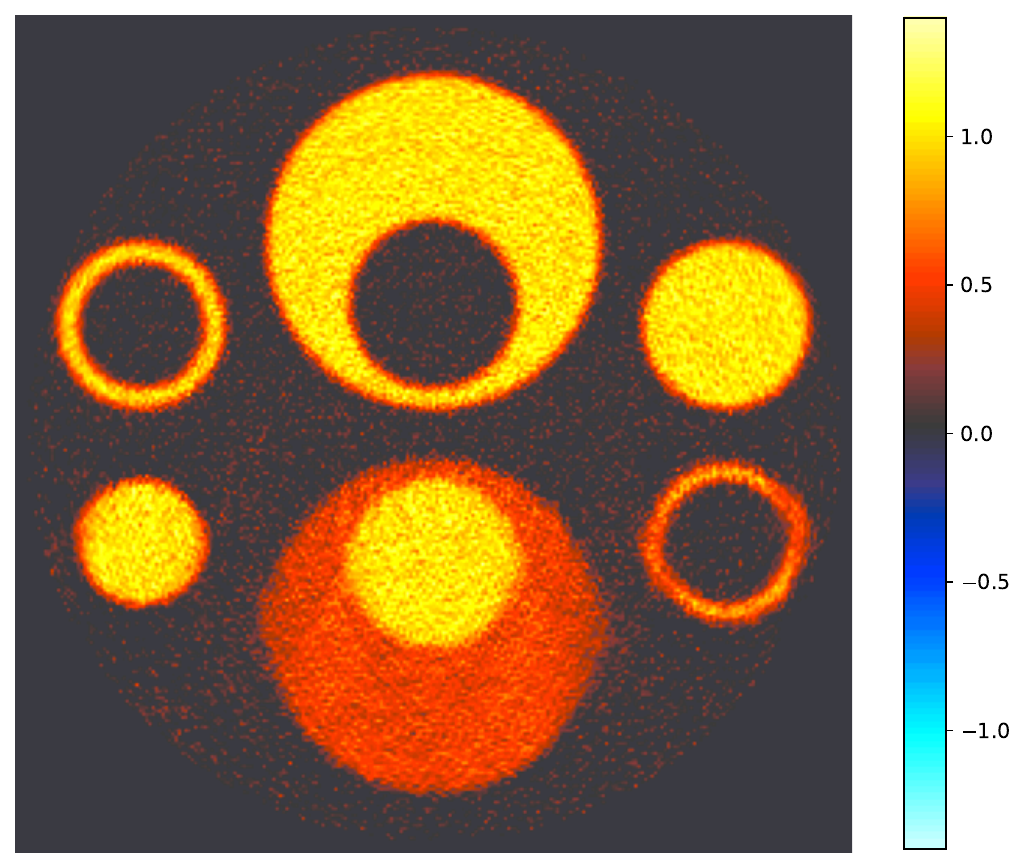}}
\subfigure[Iterative reconstruction, TV]
{\includegraphics[scale=0.4]{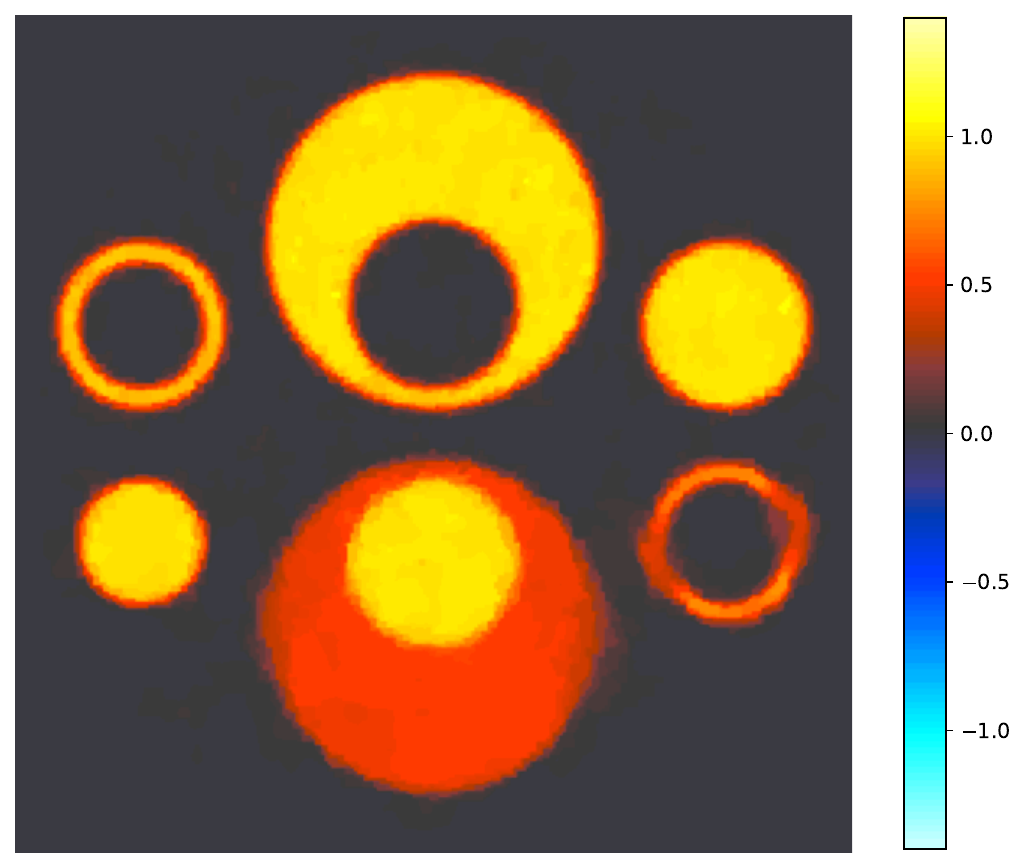}}
\subfigure[Iterative reconstruction, LPD]
{\includegraphics[scale=0.32]{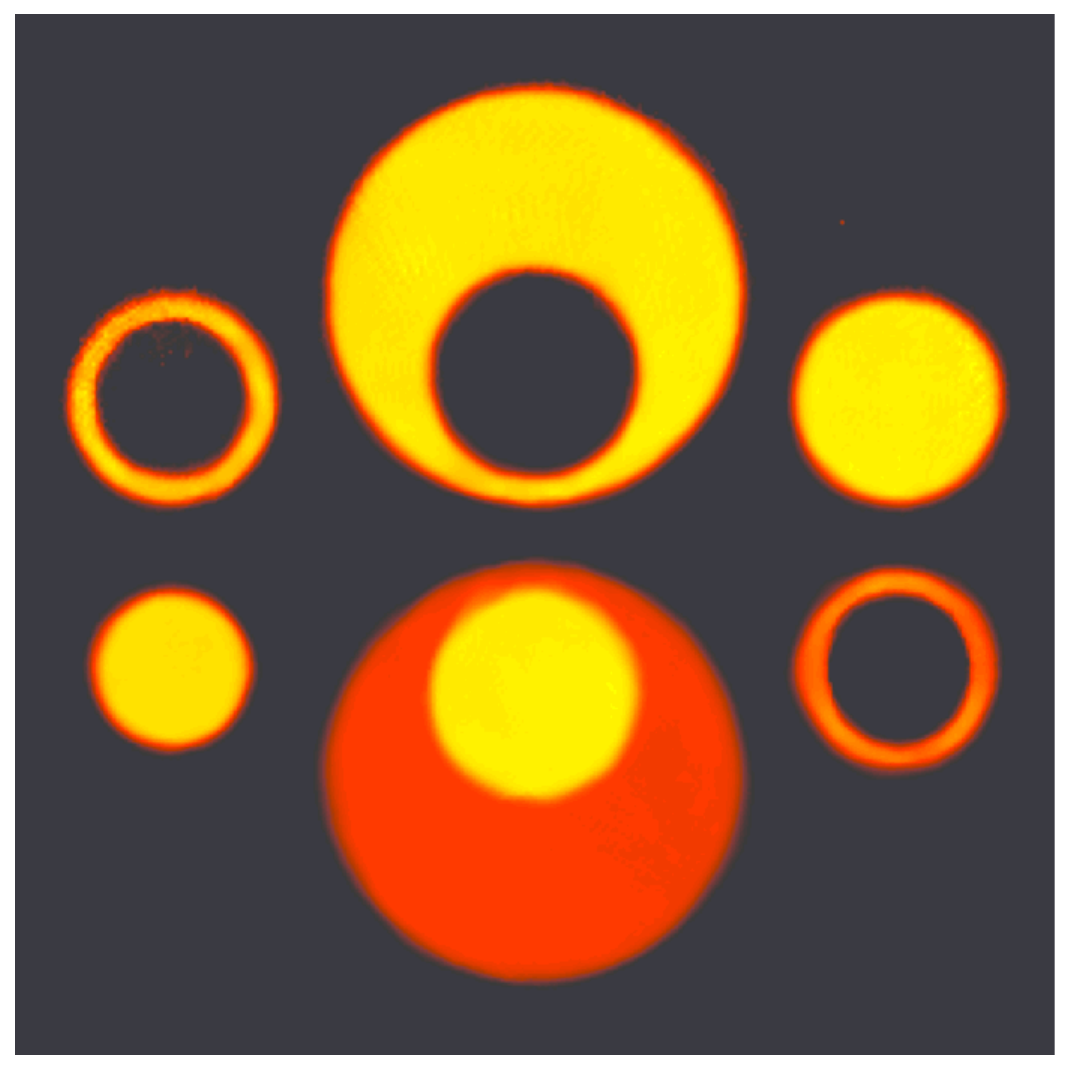}\phantom{aaaaa}}
\end{center}
\par
\vspace{-0.5cm}\caption{180 deg acquisition, incomplete data, i.e. support is
not satisfying visibility condition, data with 30\% noise}
\label{F:180full}
\end{figure}

\begin{figure}[t]
\begin{center}
\subfigure[The inverse $\mathcal{A}^{-1} g$]
{\includegraphics[scale=0.4]{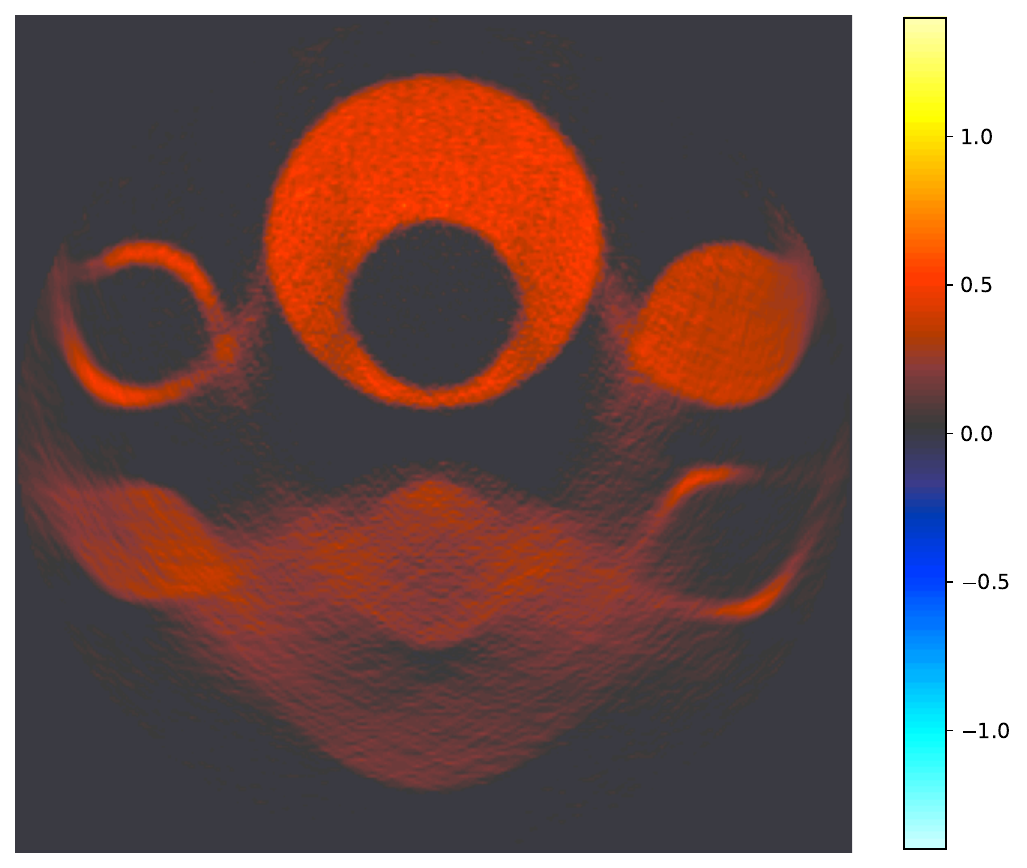}}
\subfigure[NNLS iterations]
{\includegraphics[scale=0.4]{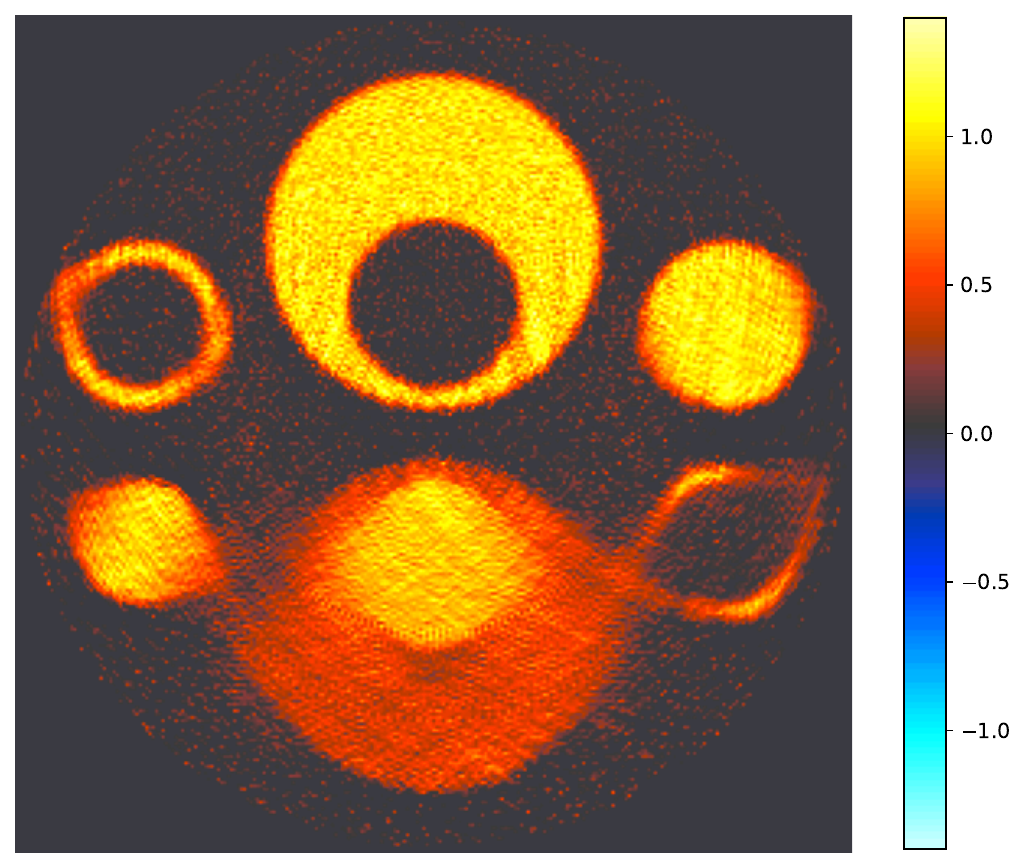}}
\subfigure[Iterative reconstruction, TV]
{\includegraphics[scale=0.4]{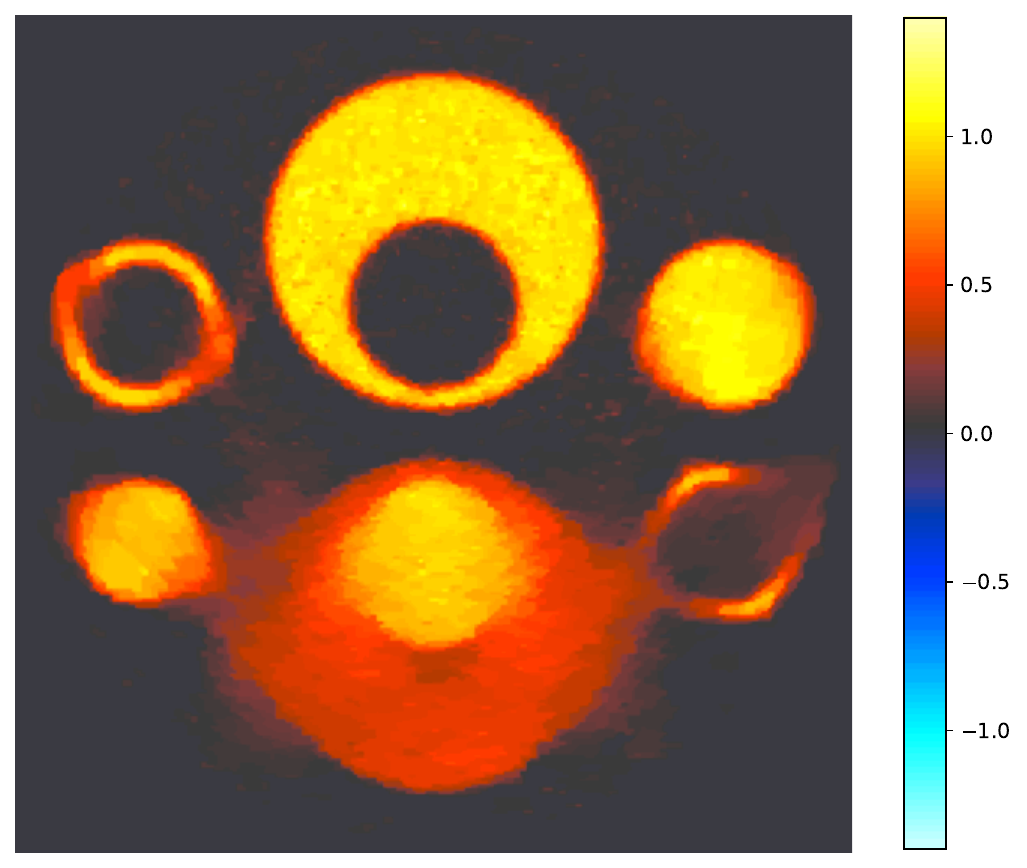}}
\subfigure[Iterative reconstruction, LPD]
{\includegraphics[scale=0.32]{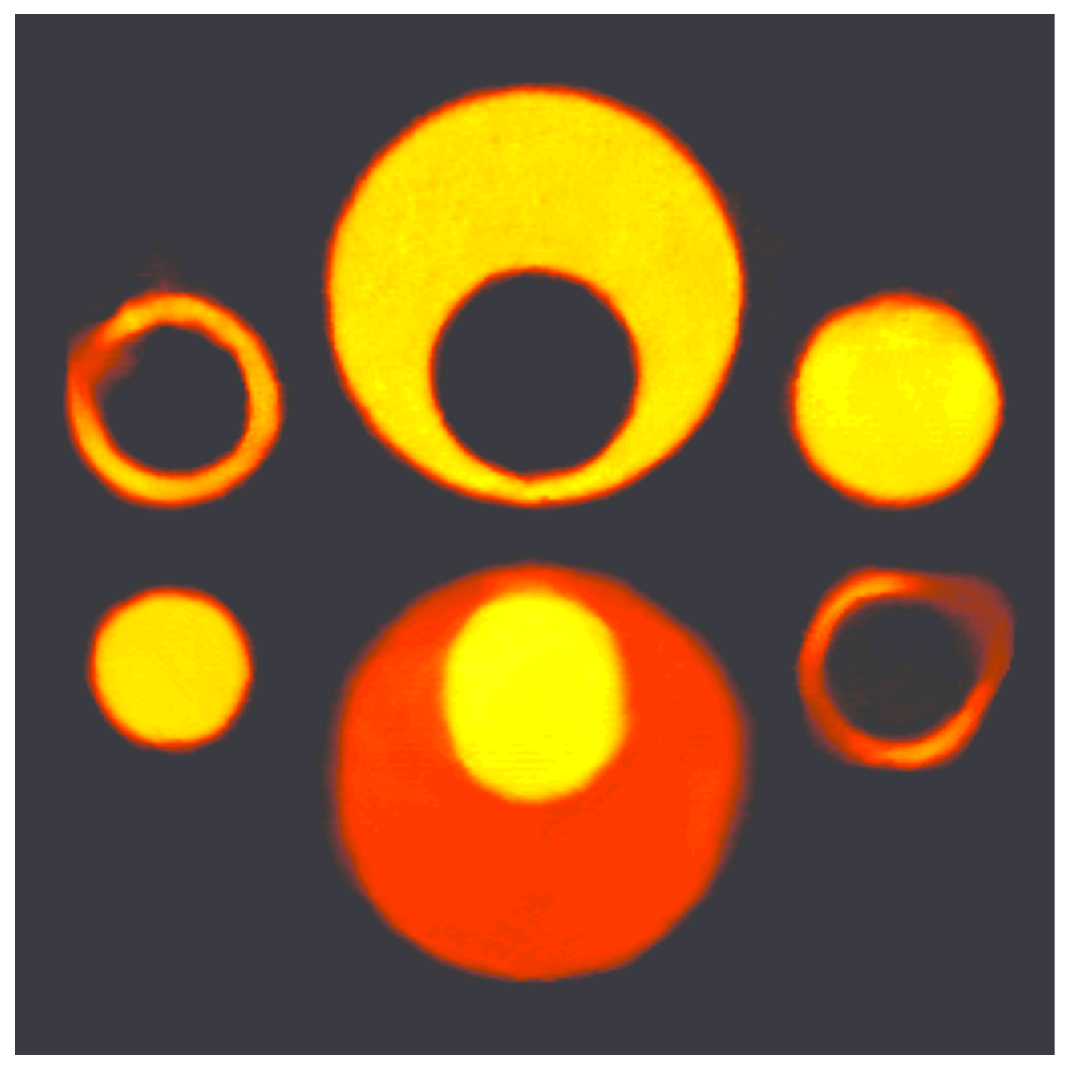}\phantom{aaaaa}}
\end{center}
\par
\vspace{-0.5cm}\caption{120 deg acquisition, incomplete data, i.e. support is
not satisfying visibility condition, with 30\% noise}
\label{F:120full}
\end{figure}

\subsection{Inverse problem with incomplete data}\label{S:half_image}

Here, we utilize our fast algorithms to solve inverse problems with incomplete
data. Namely, we consider the situation where the detectors occupy either the
upper half of the circle $S$ or a 120 degree arch of the circle, while the
function $f$ to reconstructed is supported in the disk $D.$ In this situation
a significant part of the singularities (or material interfaces) of the
function $f$ is "invisible \cite{KuKuEncycl}, and they cannot be stably
reconstructed. One can only hope to reconstruct a "visible" part of the image.
This kind of a problem calls for the use of advanced nonlinear
optimization techniques; there is hope that methods based on deep learning
can provide better images.

For these simulations, we use the phantom shown in \ref{F:360clean}(a). The
first set of images is obtained for the data $g$ measured on the upper half of
$S$, with added 30\% noise. Figure \ref{F:180full}(a) presents the image
obtained by application of $\mathcal{A}^{-1}$. It is not accurate: the
relative $L_{2}$ error is 49\% and the relative error in the $L^{\infty}$ norm is 76\%. The NNLS algorithm converges in 83 iterations
taking 3.2 sec. It produces an improved image with $L_{2}$ error reduced to
18\% and $L^{\infty}$ error equal to 62\%, see Figure \ref{F:180full}(b). The
iterative TV reconstruction produces the image shown in \ref{F:180full}(c). It
also takes 83 iterations; the error in the relative $L_{2}$ norm is 8.2\%, and
in the relative $L^{\infty}$ norm it measures 50\%. Note, however, that all
three techniques do not reconstruct the "invisible" boundaries of the
characteristic functions of circles (the boundaries with nearly horizontal
normals in the lower part of the images), and they appear blurred, as
predicted by general theory. The image reconstructed by the LPD algorithm is presented in Figure~\ref{F:180full}(d). The image is much closer to the ground truth visually; the error in the relative $L_{2}$ norm is 5\%, and
in the relative $L^{\infty}$ norm it is equal to 38\%. Training of the neural network
for the LPD technique took about 25 hours on Nvidia A2000 GPU. It involved 2 million of evaluations of
operators $\mathcal{A}$  and $\mathcal{A}^*$ (each). The benefit of using the proposed fast algorithms
for computing these operators is evident in this example.
\begin{table}[t]
\begin{tabular}
[c]{|l|r|r|c|c|c|c|}\hline
Acquisition scheme & Method & {\small Number} & Full & Time &  Error &
Error\\
 &  & of & time & per it-n & $L_{2}$ & $L_{\infty}$ \\
&  & iter-ns & (sec.) & (sec.) & (\%) & (\%)\\\hline \hline
\multirow{2}{*}{$360^\circ$, no noise } & Inverse & 1 & 1.3 (0.011) & -
& 0.22 & 0.9\\\cline{2-7}
& Reversal & 1 & 0.19 & - & 11 & 30\\\hline
\multirow{2}{*}{$360^\circ$, 30\% noise } & Inverse & 1 & 1.3
(0.011) & - & 9.9 & 30\\\cline{2-7}
& TV & 53 & 2.8 & 0.053 & 5.5 & 22\\\hline  \hline
\multirow{2}{*}{$180^\circ$ (partial), no noise } & Inverse & 1 & 1.3 (0.011) & -
& 40 & 51\\\cline{2-7}
& NNLS & 17 & 2. & 0.12 & 0.5 & 2.8\\\hline
\multirow{2}{*}{$180^\circ$ (partial), 30\% noise } & NNLS & 26 & 2.2 & 0.085 &
11 & 37\\\cline{2-7}
& TV & 74 & 3 & 0.041 & 5.2 & 26\\\hline  \hline
\multirow{5}{*}{$180^\circ$ (incomplete), 30\% noise} & Inverse & 1 & 1.3
(0.011) & - & 49 & 76\\\cline{2-7}
& NNLS & 83 & 3.2 & 0.039 & 18 & 62\\\cline{2-7}
& TV & 83 & 4. & 0.048 & 8.2 & 50\\\cline{2-7}
& LPD & 10 & 0.55 & 0.055 & 5 & 38\\\hline \hline
\multirow{5}{*}{$120^\circ$ (incomplete), 30\% noise} & Inverse & 1 & 1.3
(0.011) & - & 66 & 95\\\cline{2-7}
& NNLS & 231 & 6 & 0.026 & 26 & 79\\\cline{2-7}
& TV & 137 & 4.3 & 0.031 & 20 & 69\\\cline{2-7}
& LPD & 10 & 0.55 & 0.055 & 11 & 63\\\hline
\end{tabular}

\caption{Summary of numerical simulations. The numbers in parentheses in the fourth column show computation time not counting pre-computation of Bessel functions.}
\end{table}

Finally, for the Figure \ref{F:120full} we used the data restricted to the 120
arch of circle $S$ located symmetrically on the top of the circle, with 30\%
noise added. The application of $\mathcal{A}^{-1}$ produces an inaccurate
image shown in Figure \ref{F:120full}(a). Here the relative $L_{2}$ error is
66\% and the relative error in $L^{\infty}$ norm is 95\%. The forward/adjoint
iterative algorithm converges in 231 iterations, taking 6 seconds. The
corresponding image is presented in Figure \ref{F:120full}(b); the $L_{2}$
error is 26\%, and the $L^{\infty}$is 79\%. Interestingly, the iterative
algorithm with TV regularization takes only 137 iterations to converge,
requiring 4.3 seconds of computation time. The $L_{2}$ error is reduced to
26\% in this case, and $L^{\infty}$ error is 69\%. As expected, in all these
images "invisible" boundaries are blurred. Figure~\ref{F:120full}(d) shows the image
obtained by the LPD technique. The noise is visibly reduced, and the error in
the $L_{2}$ and $L^{\infty}$ norms is equal to 11\% and 63\% respectively.
Network training for the LPD method took the same amount of time as in the previous example.

The results of our numerical simulations are summarized in Table 1.

\section{Conclusions}

We have presented asymptotically fast algorithms for efficient numerical evaluation of operators $\mathcal{A}$, $\mathcal{A}^\ast$, and $\mathcal{A}^{-1}$ arising in the inverse problem of PAT/TAT with a circular data acquisition scheme.
The efficiency of our techniques has been verified in a series of numerical experiments, in conjunction with various image reconstruction approaches, ranging from a straightforward application of the inverse operator, to TV image regularization
and a deep-learning assisted method. A significant speed-up in computations provided by our algorithms is especially valuable
in the context of iterative image reconstruction and deep-learning assisted tomography, where the number of operator evaluations ranges from hundreds to millions (for network training purposes).  However, the purpose of the present paper was not to establish which approach is "better", but rather to demonstrate the efficiency and versatility of the proposed fast algorithms, and their usefulness for future research of optimization and deep learning in the problems of PAT and TAT. Finally, we have published the codes for the presented methods as open source to promote usage and dissemination.

\section*{Appendix}

Let us find a set of formulas to compute the function $v(x)$ given by the
equation (\ref{E:approx_UBP}) that will result in a fast algorithm. Starting
with the Fourier transform representation of $G(t,x)$ (equation
(\ref{E:Fourier_Green})), one obtains for the normal derivative $\frac
{\partial}{\partial n(z)}G(t,x-z)$ the following expression:

\begin{align*}
\frac{\partial}{\partial n(z)}G(t,x-z)  &  =\hat{z}(\theta)\cdot\nabla
_{z}G(t,x-z)=\hat{z}(\theta)\cdot\nabla_{z}\left[  \mathcal{F}_{2D}
^{-1}\left(  \frac{\sin|\xi|t}{|\xi|}\right)  \right]  (t,x-z)\\
&  =\frac{1}{2\pi}\hat{z}(\theta)\cdot\nabla_{z}\left(  \int
\limits_{\mathbb{R}^{2}}\frac{\sin|\xi|t}{|\xi|}\,e^{i\xi\cdot(x-z)}
\,d\xi\right)  =\frac{i}{2\pi}\int\limits_{\mathbb{R}^{2}}(\hat{z}
(\theta)\cdot\xi)\frac{\sin|\xi|t}{|\xi|}\,e^{i\xi\cdot(x-z)}\,d\xi\\
&  =\frac{i}{2\pi}\int\limits_{\mathbb{R}^{2}}\cos(\theta-\varphi)\sin
(|\xi|t)e^{i\xi\cdot(x-z)}\,d\xi.
\end{align*}
Now expression (\ref{E:approx_UBP}) for $v(x)$ can be transformed as follows:
\begin{align*}
v(x)  &  =2\int\limits_{0}^{T}\int\limits_{0}^{2\pi}g(t,\hat{z}(\theta
))\left(  \frac{i}{2\pi}\int\limits_{\mathbb{R}^{2}}\cos(\theta-\varphi
)\sin(|\xi|t)e^{i\xi\cdot(x-z)}\,d\xi\right)  \,d\theta\,dt\\
&  =\frac{i}{\pi}\int\limits_{\mathbb{R}^{2}}\left[  \int\limits_{0}
^{T}\left(  \int\limits_{0}^{2\pi}g(t,\hat{z}(\theta))\cos(\theta-\varphi
)\sin(|\xi|t)\,e^{-i\xi\cdot z}\,d\theta\right)  \,dt\right]  e^{i\xi\cdot
x}\,d\xi=\left[  \mathcal{F}_{2D}^{-1}\hat{v}\right]  (x),
\end{align*}
where
\[
\hat{v}(\xi)=2i\int\limits_{0}^{T}\left(  \int\limits_{0}^{2\pi}g(t,\hat
{z}(\theta))\cos(\theta-\varphi)\sin(\lambda t)\,e^{-i\xi\cdot z}
\,d\theta\right)  \,dt
\]
Use the Fourier series (\ref{E:FourSerForg}) for $g$ to simplify the inner
integral in the above equation:
\begin{align*}
\int\limits_{0}^{2\pi}\left[  \sum_{k=-\infty}^{\infty}g_{k}(t)e^{ik\theta
}\right]  \cos(\theta-\varphi)\sin(\lambda t)\,e^{-i\xi\cdot z}\,d\theta &
=\sum_{k=-\infty}^{\infty}g_{k}(t)\sin(\lambda t)\int\limits_{0}^{2\pi
}e^{ik\theta}\cos(\theta-\varphi)\,e^{-i\xi\cdot z}\,d\theta\\
&  =\sum_{k=-\infty}^{\infty}g_{k}(t)\sin(\lambda t)I_{k}(\varphi,\lambda),
\end{align*}
where we denote by $I_{k}(\varphi,\lambda)$ the following integrals:
\begin{align*}
I_{k}(\varphi,\lambda)  &  \equiv\int\limits_{0}^{2\pi}e^{ik\theta}\cos
(\theta-\varphi)\,e^{-i\xi\cdot z}\,d\theta=\frac{1}{2}\int\limits_{0}^{2\pi
}e^{ik\theta}(e^{i(\theta-\varphi)}+e^{-i(\theta-\varphi)})\,e^{-i\lambda
\cos(\theta-\varphi)}\,d\theta\\
&  =\frac{1}{2}e^{ik\varphi}\left[  \int\limits_{0}^{2\pi}e^{i(k+1)\theta
}\,e^{-i\lambda\cos\theta}\,d\theta+\int\limits_{0}^{2\pi}e^{i(k-1)\theta
}\,e^{-i\lambda\cos\theta}\,d\theta\right]  .
\end{align*}
Using the Jacobi-Anger expansion (\ref{E:Anger}) one obtains:
\[
\int\limits_{0}^{2\pi}e^{i(k\pm1)\theta}\,e^{-i\lambda\cos\theta}
\,d\theta=\int\limits_{0}^{2\pi}e^{i(k\pm1)\theta}\left(  \sum
\limits_{n=-\infty}^{\infty}(-i)^{|n|}J_{|n|}(\lambda)e^{in\theta}\right)
\,d\theta=2\pi(-i)^{|k\pm1|}J_{|k\pm1|}(\lambda).
\]
Now
\begin{equation}
I_{k}(\varphi,\lambda)=\pi e^{ik\varphi}[(-i)^{|k+1|}J_{|k+1|}(\lambda
)+(-i)^{|k-1|}J_{|k-1|}(\lambda)]. \label{E:eq1A}
\end{equation}
By taking into account well-known properies of the Bessel functions,
\[
J_{1}(\lambda)=-J_{0}^{^{\prime}}(\lambda),\qquad\frac{1}{2}(J_{m+1}
(\lambda)-J_{m-1}(\lambda))=-J_{m}^{^{\prime}}(\lambda),\quad m\in\mathbb{Z},
\]
formula (\ref{E:eq1A}) simplifies to
\[
I_{k}(\varphi,\lambda)=2\pi ie^{ik\varphi}(-i)^{|k|}J_{|k|}^{^{\prime}
}(\lambda).
\]
We now have
\begin{align*}
& \hat{v}(\xi) = \\
&  =2i\int\limits_{0}^{T}\left(  \sum_{k=-\infty}^{\infty}
g_{k}(t)\sin(\lambda t)I_{k}(\varphi,\lambda)\right)  \,dt=-4\pi
\int\limits_{0}^{T}\left(  \sum_{k=-\infty}^{\infty}g_{k}(t)\sin(\lambda
t)e^{ik\varphi}(-i)^{|k|}J_{|k|}^{^{\prime}}(\lambda)\right)  \,dt\\
&  =-4\pi\sum_{k=-\infty}^{\infty}e^{ik\varphi}(-i)^{|k|}J_{|k|}^{^{\prime}
}(\lambda)\left(  \int\limits_{0}^{T}g_{k}(t)\sin(\lambda t)\,dt\right)
\end{align*}
or
\[
\hat{v}(\xi)=\sum_{k=-\infty}^{\infty}e^{ik\varphi}\hat{v}_{k}(\lambda
),\qquad\hat{v}_{k}(\lambda)\equiv-4\pi(-i)^{|k|}J_{|k|}^{\prime}(\lambda
)\int\limits_{0}^{T}g_{k}(t)\sin(\lambda t)\,dt,
\]
which coincides with equations\ (\ref{E:Four_ser_v}), (\ref{E:Four_coef_v}).

\bigskip

\bigskip

\bigskip

\bigskip

\section*{Acknowledgments}

We would like to thank Janek Gröhl for providing Figure \ref{F:existing}.

This work is supported by the Research Council of Finland with the Flagship of Advanced Mathematics for Sensing Imaging and Modelling Project Nos. 359186, 358944, Centre of Excellence of Inverse Modelling and Imaging Project Nos. 353093, 353086, and the Academy Research Fellow Project No. 338408. The Finnish Ministry of Education and Culture's Pilot for Doctoral Programmes (Pilot project Mathematics of Sensing, Imaging and Modelling). The European Research Council (ERC) under the European Union's Horizon 2020 Research and Innovation Programme (Grant Agreement No. 101001417—QUANTOM). L.K. is partially supported by the NSF, through the Award No. NSF/DMS-2405348.

\bibliographystyle{siam}
\bibliography{bibi}

\end{document}